\documentclass{aa}  

\usepackage{graphicx}
\usepackage{txfonts}
\usepackage{color}
\usepackage{soul}
\DeclareUnicodeCharacter{2212}{\ensuremath{-}}
\DeclareRobustCommand{\masa}{\,mas\,a$^{-1}$}
\DeclareRobustCommand{\gaia}{\textit{Gaia} }

\DeclareRobustCommand{\upmask}{\texttt{UPMASK}}
\DeclareRobustCommand{\limepy}{\texttt{LIMEPY}}
\DeclareRobustCommand{\spes}{\texttt{SPES}}
\DeclareRobustCommand{\python}{\texttt{PYthon}}
\newcommand{\kms}{\,km\,s$^{-1}$}

\begin{document}

   \title{The longevity of the oldest open clusters:}

   \subtitle{Structural parameters of NGC\,188, NGC\,2420, NGC\,2425, NGC\,2682, NGC\,6791, NGC\,6819}

   \author{N. Alvarez-Baena\inst{1}\fnmsep\thanks{RECA 2021 intership programme.}
          \and
          R. Carrera  \inst{2,3}
          \and
          H. Thompson\inst{4,2}\fnmsep\thanks{Erasmus+ student.}
          \and
          L. Balaguer-Nu\~nez\inst{5,6,7}
          \and
          A. Bragaglia\inst{2}
          \and
          C. Jordi\inst{5,6,7}
          \and
          E. Silva-Villa\inst{1}
          \and
          A. Vallenari\inst{3}
          }

   \institute{Instituto de F\'{\i}sica, Universidad de Antioquia,  Calle. 67 No. 53-108, A. A. 1226, Medell\'{\i}n, Colombia\\
              \email{natalia.alvarez3@udea.edu.co}
        \and 
        INAF-Osservatorio di Astrofisica e Scienza dello Spazio di Bologna, via P. Gobetti 93/3, 40129, Bologna, Italy\\
        \email{jimenez.carrera@inaf.it}
        \and
        INAF-Osservatorio Astronomico di Padova, vicolo Osservatorio 5, 35122, Padova, Italy
        \and
             Department of Physics, University of Surrey, Guildford GU2 7XH, UK
        \and Institut d'Estudis Espacials de Catalunya (IEEC), Gran Capit\`a, 2-4, 08034 Barcelona, Spain 
 \and
 Institut de Ci\`encies del Cosmos (ICCUB), Universitat de Barcelona (UB), Martí i Franquès, 1, 08028 Barcelona, Spain
 \and
 Departament de Física Quàntica i Astrofísica (FQA), Universitat de Barcelona (UB),  Martí i Franquès, 1, 08028 Barcelona, Spain 
             }

% \abstract{}{}{}{}{} 
% 5 {} token are mandatory
 
  \abstract
   {Open clusters' dynamical evolution is driven by stellar evolution, internal dynamics and external forces, which according to dynamical simulations, will evaporate them in a timescale of about 1\,Ga. However, about 10\% of the known open clusters are older. They are special systems whose detailed properties are related to their dynamical evolution and the balance between mechanisms of cluster formation and dissolution.}
   {We investigate the spatial distribution and structural parameters of six open clusters older than 1\,Ga in order to constrain their dynamical evolution, and longevity.}
   {We identify members using \gaia EDR3 data up to a distance of 150\,pc from each cluster's center. We investigate the spatial distribution of stars inside each cluster to understand their degree of mass segregation. Finally, in order to interpret the obtained radial density profiles we reproduced them using the lowered isothermal model explorer with \python~(\limepy) and spherical potential escapers stitched (\spes).}
   {All the studied clusters seem more extended than previously reported in the literature. The spatial distributions of three of them show some structures aligned with their orbits. They may be related to the existence of extra tidal stars. In fact, we find that about 20\% of their members have enough energy to leave the systems or are already unbound. Together with their initial masses, their distances to the Galactic plane may play significant roles in their survival. We found clear evidences that the most dynamically evolved clusters do not fill their Roche volumes, appearing more concentrated than the others. Finally, we find a cusp-core dichotomy in the central regions of the studied clusters, which shows some similarities to the one observed among globular clusters.}
  {}


   \maketitle
%
%-------------------------------------------------------------------

\section{Introduction}

Open clusters are particularly helpful to understand the Galactic disc since some of their properties, such as ages or distances, are more accurately determined in comparison with other tracers. They were formed from the very earliest disc stages, as evidenced by the existence of the oldest clusters, such as Berkeley\,17 \citep[e.g.][]{bragaglia2006}. As clusters continue to form within the process of overall star formation, they reflect gradients in disc structural and chemical properties \citep[e.g.][]{carrera_pancino2011,ApogCarrera19,carrera2022occaso}. Open clusters are dynamically-bound groups formed from a dozen to several thousand members. For a given cluster, all the stars share chemo-dynamical features with a common birth time and motion. Each system is chemically homogeneous, for the majority of elements, at the level of precision of current abundance determination of 0.05\,dex \citep[e.g.][]{bovy2016,casamiquela2020,poovelil2020,patil2022}, excluding those elements (e.g., Li, C, N, Fe) whose atmosphere abundances are modified by diffusion during the stellar evolution \citep[e.g.][]{bertelli_motta2018,Hasselquist2019,charbonnel2020}. These features have motivated the use of open clusters as probes of a variety of astrophysical phenomena such as stellar evolution and nucleosynthesis, stellar interactions or star formation.

\begin{table*}
\setlength{\tabcolsep}{0.5mm}
	\centering
	\caption{Properties of the sampled clusters.}
	\label{tab:initialparameter}
	\begin{tabular}{lcccccccccccccc}
		\hline
		Cluster & RA\tablefootmark{a} & DEC\tablefootmark{a} & $\mu_{\alpha*,DR3}$\tablefootmark{b} & $\mu_{\delta,DR3}$\tablefootmark{b}  & $\varpi_{DR3}$\tablefootmark{b}  & Distance\tablefootmark{a}  & Age\tablefootmark{c} & \textit{X}\tablefootmark{a} & \textit{Y}\tablefootmark{a} & \textit{Z}\tablefootmark{a} & $R_{\rm gc}$\tablefootmark{a} & $R_{\rm 150}$\tablefootmark{d} & $N_{\rm member}$ & [Fe/H]\\
		&[$\degr$] & [$\degr$] & [\masa] & [\masa] & [mas] & [pc] & [Ga] & [pc] & [pc] & [pc] & [kpc] & [$\degr$] & [star] & [dex]\\
		\hline
  NGC\_188 & 11.798 & 85.244 & -2.3$\pm$0.1 & -1.0$\pm$0.1 & 0.51$\pm$0.05 & 1698 & 7.5$\pm$0.02 & -851 & 1319 & 646 & 9.3 & 4.6 & 1143 & -0.03$\pm$0.07\tablefootmark{e}\\
  NGC\_2420 & 114.602 & 21.575 & -1.2$\pm$0.1 & -2.1$\pm$0.1 & 0.36$\pm$0.06 & 2587 & 1.9$\pm$0.02 & -2316 & -757 & 869 & 10.7 & 3.4 & 682 & -0.22$\pm$0.03\tablefootmark{e}\\
  NGC\_2425 & 114.577 & -14.885 & -3.6$\pm$0.1 & 2.0$\pm$0.1 & 0.26$\pm$0.07 & 3576 & 2.1$\pm$0.01 & -2222 & -2794 & 205 & 10.9 & 2.4 & 455 & -0.13$\pm$0.03\tablefootmark{f}\\
  NGC\_2682 & 132.846 & 11.814 & -11.0$\pm$0.2 & -3.0$\pm$0.2 & 1.13$\pm$0.05 & 889 & 3.6$\pm$0.02 & -613 & -440 & 470 & 9.0 & 9.9 & 1164 & 0.04$\pm$0.04\tablefootmark{e}\\
  NGC\_6791 & 290.221 & 37.778 & -0.4$\pm$0.2 & -2.3$\pm$0.2 & 0.19$\pm$0.08 & 4231 & 8.4$\pm$0.04 & 1423 & 3903 & 800 & 7.9 & 1.9& 3669 & 0.15$\pm$0.14\tablefootmark{e}\\
  NGC\_6819 & 295.327 & 40.19 & -2.9$\pm$0.1 & -3.9$\pm$0.1 & 0.36$\pm$0.05 & 2765 & 2.0$\pm$0.01 & 754 & 2628 & 407 & 8.0 & 3.3& 2112 & 0.04$\pm$0.06\tablefootmark{e}\\
\hline
	\end{tabular}
 \tablefoot{\tablefoottext{a}{Values obtained from \citet{cantatgaudin2020} derived from \gaia DR2 \citep{GaiaBrown}}\tablefoottext{b}{Values recomputed using \gaia DR3 \citep{gaiadr3_content} from the membership probabilities published by \citet{cantatgaudin2020}.} \tablefoottext{c}{Values from \citet{bossini2019}.}\tablefoottext{d}{The projection on sky corresponds to a physical radius of 150\,pc for each system.}\tablefoottext{e}{From Carbajo-Hijarrubia et al. {\sl in prep.}}\tablefoottext{f}{From \citet{randich2022ges_oc}.}}
\end{table*}

During their lives, open clusters are strongly affected by stellar evolution, internal dynamics and external forces \citep[see][for a recent review]{Krumholz2019}. Very young clusters, $<$40\,Ma, suffer the so-called infant mortality. Protostellar outflows, photoionization, radiation pressure, or supernova shocks expel high speed gas that is able to evaporate the less concentrated systems \citep[e.g.][]{lada_lada2003ARA&A,bally2016ARA&A,kim2018,krumholz2018}. The dynamical evolution of surviving gas-free clusters is driven by relaxation. The stars randomly exchange energy via gravitational interactions, causing equipartition of energy between stars of different masses. This causes a mass segregated system with the most massive objects concentrated in the centre while the lower mass stars migrate to the outskirts, forming a halo. Some of these stars acquire enough velocity to escape from the system, resulting in its gradual evaporation \citep[e.g.][]{pang2021}. This dissolution is amplified by the forces acting on these systems as they orbit in the Galaxy, such as encounters with giant molecular clouds or passes through the disc \citep[e.g.][]{binney_tremaine2008,gieles2006}. Therefore, according to dynamical simulations a typical cluster, with $10^4$\,M$_{\odot}$, will evaporate in a timescale of $\sim$1\,Ga \citep[e.g.][]{baumgardt2003,lamers2005,bastian2008}.

Between 8 and 10\% of the more than 6,800 known and candidate open clusters have ages older than 1\,Ga \citep{cantatgaudin2020,hunt2023}. The properties of this older population are related to the dynamical evolution of clusters and the balance between mechanisms of cluster formation and dissolution \citep{friel2013}. These clusters are preferentially found at galactocentric distances larger than 6\,kpc, and at larger heights from the Galactic plane. The majority are found close to their maximum excursion from the plane, where they spend most of the time, away from the disc disruptive influence.

The oldest systems are typically larger than intermediate-age clusters, 50\,Ma to 1\,Ga \citep[e.g.][]{friel2013}. This, together with the preferential location in the outer disc, could be interpreted as larger clusters easily survive at larger distances. Nevertheless, this could be due to a selection effect, since small clusters are more difficult to detect. In spite of this, the longevity of the oldest open clusters is still not well understood.

The goal of this paper is to determine structural parameters and study the spatial distribution of stellar populations inside six of the oldest open clusters: NGC\,188, NGC\,2420, NGC\,2425, NGC\,2682, NGC\,6791, and NGC\,6819. This paper is organised as follows. The observational material used and astrometric membership probability determination are described in Sect.~\ref{sec:membership}. The radial density profiles and their interpretation based on dynamical models are presented in Sect.~\ref{sec:density_profiles}. The  segregation in mass of studied clusters is investigated in Sect.~\ref{sec:mass_segregation}. Several physical parameters such as Jacobi radius, half-mass relaxation time or initial mass are estimated in Sect.~\ref{sec:analysis}.
The results are discussed in the context of the open cluster dynamical evolution in Sect.~\ref{sec:discussion}. Finally, the main conclusions of this study are listed in Sect.~\ref{sec:conclusions}. 

\section{Membership determination}\label{sec:membership}

\begin{figure*}
	\includegraphics[width=\textwidth]{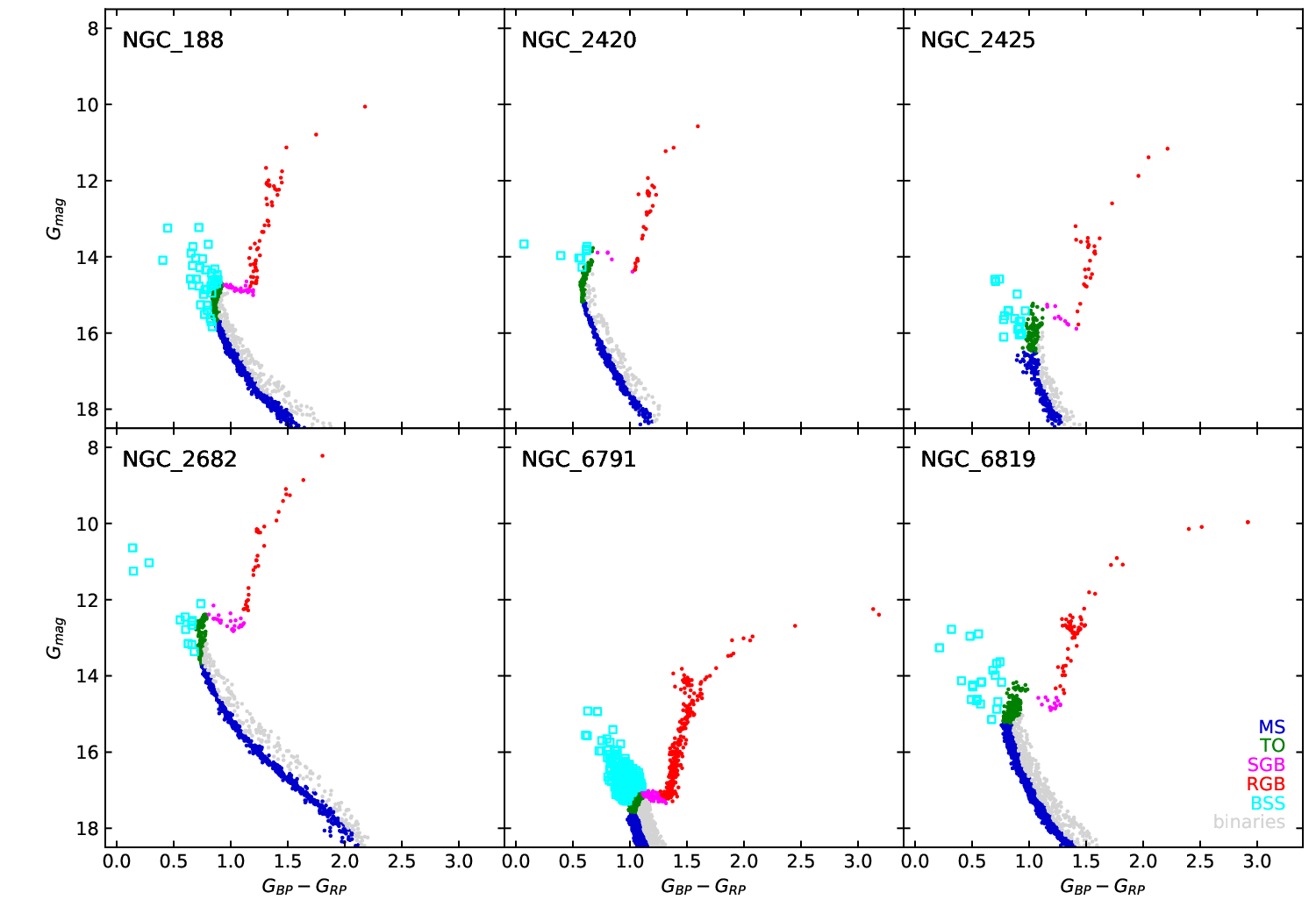}
    \caption{\gaia EDR3 colour–magnitude diagrams of the studied clusters for stars with high membership probabilities, $p\geq$0.4 (see text for details). The different populations are plotted in different colours: RGB/RC (red), SGB (magenta), TO (green), MS (blue), candidate BSS (cyan squares), and candidate binaries (light grey).}
    \label{fig:CMDs}
\end{figure*}

Our analysis is based on the positions ($\alpha$, $\delta$), proper motions ($\mu_{\alpha*}$, $\mu_{\delta}$) and parallaxes ($\varpi$) provided by \gaia EDR3 \citep{gaia_edr3,gaia_edr3astrometric} alongside magnitudes in three photometric bands \textit{G, $G_{\rm BP}$} and \textit{$G_{\rm RP}$} \citep{gaia_edr3photmetric}. We applied the \textit{G}-band corrections recommended by the \gaia team\footnote{\url{https://www.cosmos.esa.int/web/gaia/edr3-known-issues}}. Note that these are the most updated values for astrometric and photometric magnitudes provided by \gaia since DR3 only propagated the values already released in EDR3 for them \citep{gaiadr3_content}. We performed our analysis in a radius of 150\,pc from each cluster centre, as determined by \citet{cantatgaudin2020} and listed in Table~\ref{tab:initialparameter}. This value was selected because it was larger than the known cluster sizes at that time. In order to reduce the size of the sample, we applied weak constraints in proper motions and parallaxes, discarding stars outside five times the uncertainties for proper motions and parallaxes, and centred on their average values, again using the results reported by \citet[][see Table~\ref{tab:initialparameter}]{cantatgaudin2020}. Moreover, we limited our sample to stars brighter than \textit{G}=18.5\,mag to ensure a good completeness of the sample, with reasonable average uncertainties in proper motions and parallaxes. The central regions of each cluster were used to constrain their sequences in the colour-magnitude diagram, removing those objects which were far away from the cluster sequence or the position of the blue straggler stars.

For each star in these initial samples, the probability of belonging to each cluster was determined from its proper motions and parallaxes using \upmask~\citep[Unsupervised Photometric Membership Assignment in Stellar clusters,][]{upmask}. This tool, originally developed to assign membership probabilities from the photometric data, was adapted to do so by using astrometric data \citep{CG18UM, pyupmask2021}. \upmask~ uses a \textsc{k}-means clustering algorithm assuming that the member stars are closely clustered together in $\mu_{\alpha*}$, $\mu_{\delta}$, and $\varpi$ 3D space \citep[see][for a full discussion of this]{CG18UM}. Initially, we planned to follow the same procedure used by \citet{m67paper} for NGC\,2682 (M67). This procedure works well at large radii for clusters with average proper motion and parallax values statistically different from the average values of field stars, such as NGC\,2682. However, in the cases where this does not happen, \upmask~ works well at short radii where the clusters' stars dominates. If a large radius is used, \upmask~ assigns lower membership probabilities even to the central members in comparison with the values obtained using a short radius. 

To overcome this issue, the initial sample containing stars within the 150\,pc radius was split into multiple data sets containing the objects within increasing radius values. The initial radius was selected as the one which contains 50\% of the cluster members, as determined by \citet{cantatgaudin2020}. If the sample has a projection on sky larger than 3\fdg0 the data sets would have radii values increasing in steps of 0\fdg5. Alternatively, if the radius of the sample is less than 3\fdg0 the radius for each data set would increase by 0\fdg2 each time. \upmask~ was run in each data set independently. This means that the objects in the central region of each cluster have multiple membership probabilities determinations, while the stars in the outermost radius have only one. For those objects with multiple determinations, we simply assumed as membership probability the maximum value obtained, which typically is derived in the innermost radii in which this object is sampled. This provided a more exhaustive view of the centre of the cluster, whilst also keeping the maximum amount of members towards the outskirts. Finally, we considered as cluster members those objects with membership probability, \textit{p}, greater or equal than 0.4 in this final merged catalogue \citep[see][for details]{Soub18, ApogCarrera19}. The impact of this selection in our results are discussed in Sect.~\ref{sect:pthreshold}.

In the case of NGC\,2682 we found a total of 1\,170 objects with $p\geq$0.4 using the \gaia EDR3 data. This number is slightly higher than the number of stars found by \citet{m67paper} of 1\,149 from \gaia DR2 data. Not all the objects in the \citet{m67paper} sample were recovered, and instead other stars appeared with high membership probabilities. This is explained by the change in the proper motions, parallaxes, and above all, in their related uncertainties between \gaia DR2 used by \citet{m67paper} and EDR3 used here. Moreover, there is no preferential spatial location for both the new recovered and the discarded stars. Also, this issue only affects a small fraction of objects, $\sim$5\%. Therefore, this ensures that the procedures adopted here and by \citet{m67paper} are equivalent.

In the case of NGC\,6819, \citet{cantat_gaudin2018} reported 1\,715 objects in a radius of 0\fdg45 with $p\geq$0.4 and \textit{G}$\leq$18.5\,mag. We find only $\sim$1\,300 stars brighter than \textit{G}=18.5\,mag and with $p\geq$0.4 if we run \upmask~in the whole 3\fdg3 radius (150\,pc at the distance of this cluster). With the procedure adopted here of performing the analysis in different increasing radii, we recover 1\,785 objects with \textit{G}$\leq$18.5\,mag and $p\geq$0.4. All these numbers were obtained before applying constraints in the positions of the stars in the colour-magnitude diagram.

The limited capabilities of \gaia for observing dense areas, such as the centre of the clusters, can affect the completeness of our analysis. According to \citet{fabricius2021} the completeness of the sample is reduced for stars fainter than G$\sim$19\,mag for regions with densities around 10$^5$\,stars\, deg$^{-2}$. Taking into account the distance of the clusters in our sample, only the faintest objects in their very central regions may be affected by this effect, except for NGC\,2682, which is close enough to avoid problems with the crowding.

\gaia DR3 provides radial velocities for stars with \textit{G}$<$14\,mag \citep{gaiadr3_content,gaiadr3_rv}. We used this information to evaluate the contamination, at least within the brightest stars, by objects with discrepant radial velocities. Owing to the large uncertainties of the \gaia DR3 radial velocities at the faint end, we only consider as real non cluster members those stars which very discrepant radial velocities, larger than 15\kms, from the average value provided by \citet{tarricq2021} for each cluster. With this procedure, we remove stars only on four clusters. We removed two objects from a total of 26 stars with high astrometric memberships, in the case of NGC\,2425. Five objects were rejected in NGC\,2682 from a sample of 487 objects. This filter has a higher impact in NGC\,2420 and NGC\,6819 where 10 and 11 objects are discarded from the radial velocity sample of 69 and 124 stars, respectively. 

The total number of objects for each cluster after taking into account the position of the stars in the colour magnitude diagrams and their radial velocities are listed in Table~\ref{tab:initialparameter}. For each cluster, the selected members are manually separated into different groups as a function of their evolutionary stage from the position in the colour-magnitude diagrams. These groups are main-sequence (MS), turn-off (TO), sub-giant branch (SGB), red giant branch (RGB) which includes also the red clump (RC), candidate blue straggler stars (BSS) and candidate binaries. They are plotted with different colours in Fig.~\ref{fig:CMDs}. We include blue straggler stars and binaries in spite of the well known limitations of \upmask~ to assign high membership probabilities to these objects \citep[see][for a detailed discussion]{m67paper}.

\section{Radial density profiles}\label{sec:density_profiles}

In order to investigate the spatial distributions of the stellar populations inside the studied clusters, it is necessary to remove projection effects, which are important in cases such as NGC\,188 due to its location near the North celestial pole. For this purpose, we computed projected Cartesian coordinates, \textit{x} and \textit{y}, using the Eq.~2 by \citet{gaia_helmi2018}. In our case, we selected as origin, ($\alpha_0$, $\delta_0$), the cluster centre listed in Table~\ref{tab:initialparameter} taken from \citet{cantatgaudin2020}. In this system, the \textit{x}-axis is antiparallel to the right ascension axis, and the \textit{y}-axis parallel to the declination axis. We use the distances to each cluster reported by \citet{cantatgaudin2020} and listed in Table~\ref{tab:initialparameter} to convert \textit{x} and \textit{y} coordinates in parsecs. Finally, the radial distance for each star was obtained from the Cartesian coordinates defined above as $r=\sqrt{x^2+y^2}$.

\begin{figure*}
   \centering
   \includegraphics[width=\textwidth]{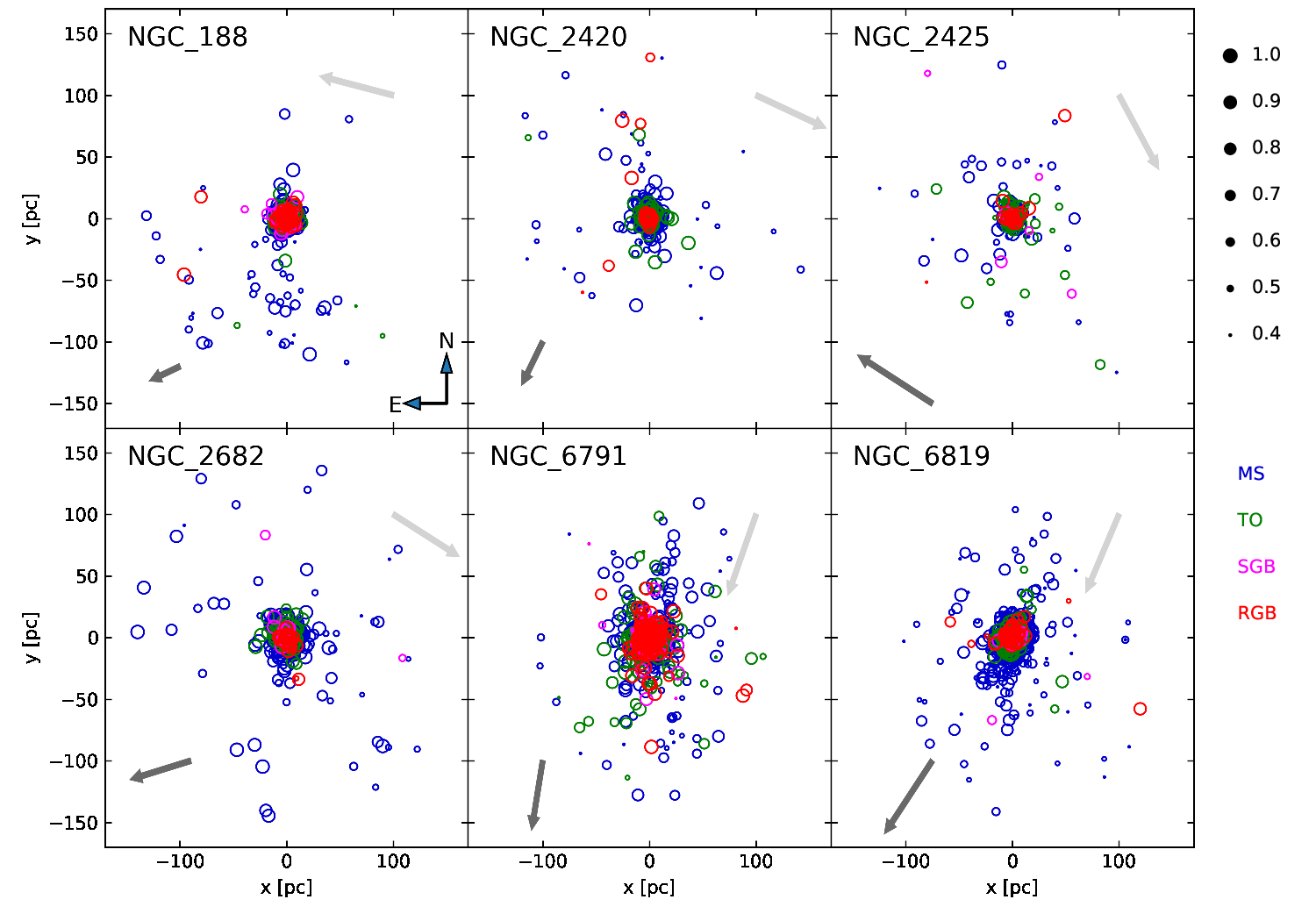}
   \caption{Spatial distribution in Cartesian coordinates (\textit{x}, \textit{y}) of the studied cluster. The size of the points is related to the astrometric membership probability.  The different populations are colour coded as in Fig.~\ref{fig:CMDs} excluding the binaries and BSS for clarity. The dark gray arrows in the bottom left corner show the direction of motion and are proportional to the velocity in each axis. The light gray arrows in the top right corner show the direction to the Galactic centre.}
   \label{fig:spt_dist}%
\end{figure*}

The spatial distributions of the studied clusters projected in the sky in Cartesian \textit{x} and \textit{y} coordinates are shown in Fig.~\ref{fig:spt_dist}, where the size of the points is related to their membership probabilities. The first noticeable feature is the tail towards the south-east of NGC\,188, almost on the direction of motion marked by the arrow. Although without a clear tail, NGC\,2420 and NGC\,2425 show non-isotropic distributions in their outskirts. In the case of NGC\,2425 it seems that it is aligned with the orbit. The other three clusters have a more regular distribution, but in the case of NGC\,6819 the central region seems to be elongated in the SE to NW direction, again almost aligned with the movement, as already reported by \citet{kamann2019}

The radial density profile is the basic tool used to investigate the spatial distribution of stellar systems, such as open clusters, providing clues about their dynamical evolution. For the studied systems, we calculated the mean stellar surface density of objects within concentric rings as $\rho_i=N_{i}/\pi(R^2_{i+1}-R^2_i)$, where $N_i$ is the number of stars within the i-th ring with an inner radius of $R_i$ and an outer radius of $R_{i+1}$. The density uncertainty in each ring was estimated assuming Poisson statistics.  

%\begin{figure*}
%   \centering
%   \includegraphics[width=\textwidth]{fig_rdprofiles_nofits.eps}
%   \caption{Radial density profiles of the studied clusters.}
%   \label{fig:rdp}
%\end{figure*}

\begin{figure*}[h!]
   \centering
   \includegraphics[width=\textwidth]{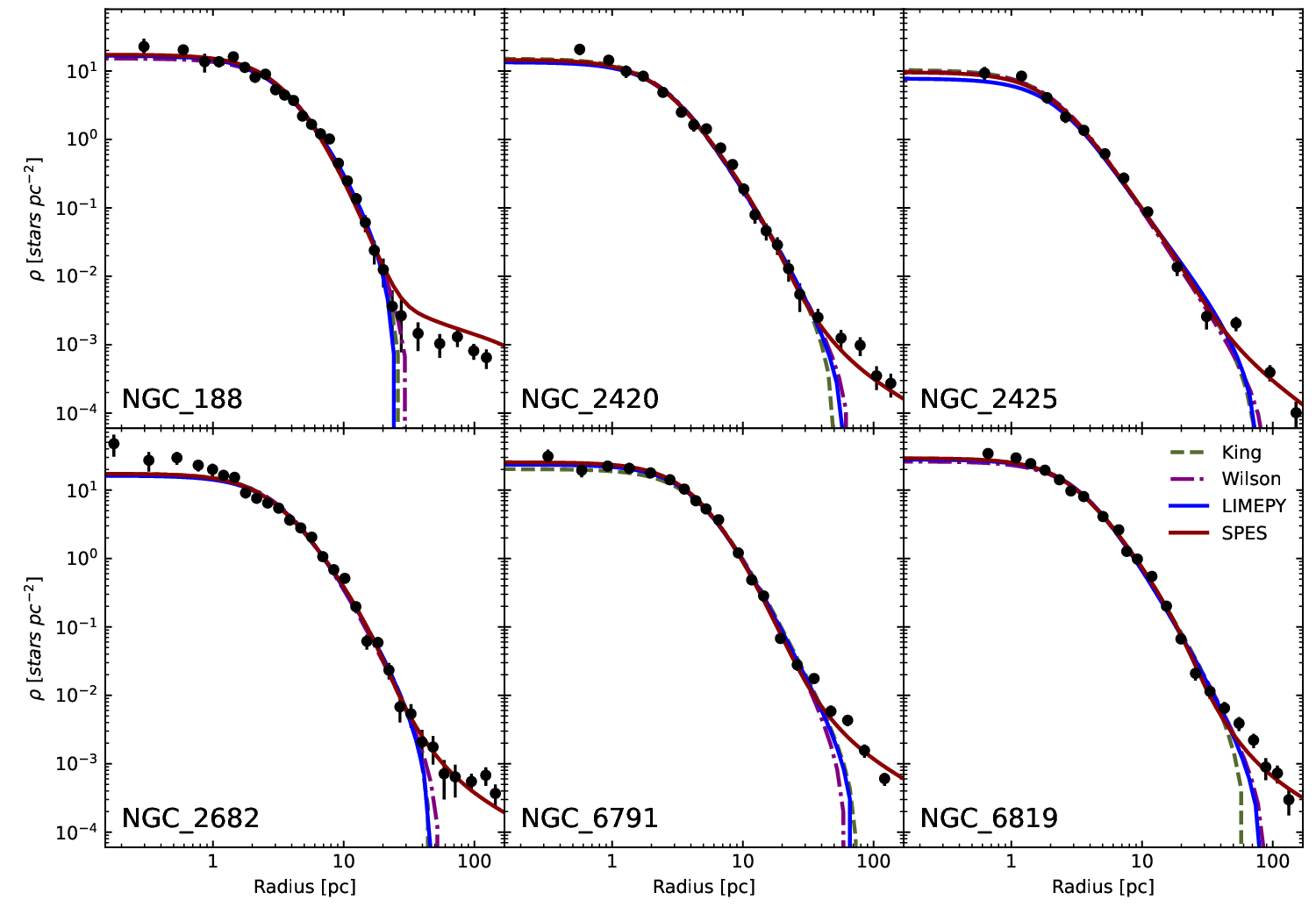}
   \caption{Radial density profiles of the studied clusters. Overplotted, the best fits King (green dashed lines), Wilson (purple dot-dashed lines), \limepy~(blue solid lines), and \spes~(red solid lines) models.}
   \label{fig:model_fits}
\end{figure*}

The obtained radial density profiles are shown in Fig.~\ref{fig:model_fits}. Only two of the studied clusters, NGC\,188 and NGC\,2682, show some hints of flattening in the outskirts. In contrast, for the other four systems, with a change of slope, the radial density profiles continue to fall until 150\,pc, which may imply that we did not reach the end of the cluster.

There are also some differences in the core regions. While some clusters show clear flat cores, such as NGC\,6791, other show a cusp profile, such as NGC\,2682 or in a lesser degree NGC\,2420. In the case of NGC\,2425, it seems that the core is smaller than in the other systems and its central region has been sampled with only two rings. The dichotomy of flat and cusp cores is well known in Galactic globular clusters \citep[e.g.][]{djorgovski1986,trager1995}. We will discuss this issue in detail in Sect.~\ref{sec:discussion}.

\subsection{LIMEPY models}

The most direct way of interpreting the observed radial density profiles is by comparison with the prediction of dynamical models. It is widely stated in the literature that in spite of the irregular appearance of open clusters, their radial density profiles are well reproduced by isothermal and spherical King models \citep{king1966}. The exception are the outermost external regions, which require an additional power-law decrease term \citep[e.g.][]{m67paper}.

\citet{davoust1977} showed that the widely used King and Wilson \citep[in the non-rotating and isotropic limit,][]{wilson1975} models are particular cases of a more general family of models. These were extended to a more general class of (isotropic) lowered isothermal models by \citet{Gomez-Leyton2014}. \citet{gieles_limepy} further expanded them by introducing parametrised prescriptions for the energy truncation, related to the edge of the cluster, and for the amount of radially biased pressure anisotropy, which determines the size of the isotropic cores. They introduced the Lowered Isothermal Model Explorer in \python~ (\limepy).\footnote{
\limepy~is available from \url{https://github.com/mgieles/limepy.}} These models are particularly suited to describe the phase-space density of stars in tidally limited, mass-segregated stellar clusters in all stages of their life-cycle.

To identify one model within the \limepy~family, it is necessary to specify the values of five parameters: the central dimensionless potential, $W_0$; the anisotropy radius, $r_{\rm a}$; the truncation parameter, $g$; the total mass of the system, $M_{\rm cl}$; and the half-mass radius, $r_{\rm h}$. The central dimensionless potential, $W_0$, is used as a boundary condition to solve the Poisson equation, and it determines the shape of the radial profiles of some relevant quantities. The anisotropy radius is related to the amount of anisotropy present in the system: the smaller it is, the more anisotropic is the model. The truncation parameter sets the sharpness of the truncation in energy: the larger it is, the more extended the models are, and the less abrupt the truncation is. When considering the isotropic version of the models, $g=0$, corresponds to the \citet{woolley1954} models, $g=1$ to the \citet{king1966} models, and $g=2$ to the non-rotating \citet{wilson1975} models. We added a sixth parameter, as suggested by \citet{zocchi2016}, to account for the background density, $\rho_{\rm bg}$.

\subsection{SPES models}

As mentioned above, \limepy~models are not able to reproduce the outermost radii sampled in each cluster. The \limepy~models provide a more elaborate description of stars near the truncation, but do not include the effect of the Galactic tidal potential, unlike other models \citep[e.g.][]{varri2009}. The tidal field makes the potential in which the stars move anisotropic, and it slows down the escape of stars \citep{fukushige2000,baumgardt2001}, because escape is limited to narrow apertures around the Lagrangian points. The result is the existence of a stellar population, known as potential escapers, which is energetically unbound, but have not yet escaped because they have not reached the Lagrangian points \citep[e.g.][]{daniel2017}. These objects are the responsible for an elevation of the density and velocity dispersion near the Jacobi radius \citep{kupper2010,claydon2017}. The presence of potential escapers in globular clusters have been suggested as the responsible for the peculiarities observed in their outskirts. 

These spherical potential escapers stitched models \citep[hereafter \spes~models,][]{spes} have an energy truncation similar to the \limepy~models discussed above, with the fundamental difference that the density of stars at the truncation energy can be non-zero. More importantly, the models include stars above the escape energy, with an isothermal distribution function that continuously and smoothly connects to the bound stars.

Apart from $W_0$, $M_{\rm cl}$, and $r_{\rm h}$, the \spes~models depend on two additional parameters, $B$ and $\eta$. The value of $B$ can be 0$\leq B \leq$1, where $B=$1 implies that there are no potential escapers. The parameter $\eta$ is the ratio of the velocity dispersion of the potential escapers over the velocity scale, and it can have values 0$\leq\eta\leq$1. For $\eta=$0 there are no potential escapers. For fixed $B$, the fraction of potential escapers correlates with $\eta$. For a fixed $\eta$, the fraction of potential escapers anticorrelates with $B$, when $B$ is close to 1. For small values of $B$, the fraction of potential escapers is approximately constant or correlates slightly with $B$ at $\eta$ constant. Finally, in the presence of potential escapers the \spes~models are not continuous at the tidal radius, $r_{\rm  t}$, but they have the ability to be solved (continuously and smoothly) beyond $r_{\rm  t}$ to mimic the effect of escaping stars \citep[see][for a detailed discussion]{spes}.

\subsection{Model fits}\label{sect:model_fit}

In order to fit the observed radial density profiles with both \limepy~and \spes~models, we use a Bayesian approach following \citet{zocchi2016} to determine the posterior probability distribution of the input parameters. For \limepy~we choose uniform priors over the following ranges: 0.8$<W_0<$15, 0.2$< g <$2.1, 0.1$< M_{\rm cl} <$10$^6$\,M$_{\odot}$,0.2$<r_{\rm h}<$30\,pc, $−$1$ < \log r_{\rm a} < $3, and $−8<\log\rho_{\rm bg}<$-1. In the case of the \spes~models together with the values of $W_0$, $M_{\rm cl}$, and $r_{\rm h}$ values described above, we select: 0$\leq B\leq$1 and 0$\leq\eta\leq$1. We use a Markov chain Monte Carlo fitting technique to explore the parameter space and to efficiently sample the posterior probability distribution for the parameters above from the \texttt{LMFIT} \python~ implementation\footnote{\texttt{LMFIT} is available at \url{https://lmfit.github.io/lmfit-py/}}. 

We consider the widely used King, $g$=1, and the isotropic and non-rotating Wilson models, $g$=2. They provide a fairly simple description of cluster morphology, with their shape entirely determined by the dimensionless central potential $W_0$. In this case, high values of $W_0$ implies more concentrated models. The third case is the isotropic, single-mass \limepy~models which fit simultaneously $W_0$ and the truncation parameter $g$. Finally, we fit the \spes~models. In each case, we performed 500 realisations for each cluster.  

\subsection{Results} \label{subsec:results}

\begin{table*}
\setlength{\tabcolsep}{1.5mm}
	\centering
	\caption{Best-fitting parameters of King, g=1, and Wilson, g=2, \limepy~models}
	\label{tab:limepy_king}
	\begin{tabular}{lcccccccccc}
		\hline
            Cluster & \multicolumn{5}{c}{King (g=1)}& \multicolumn{5}{c}{Wilson (g=2)}\\
		& $W_0$ & $M_{\rm cl}$ & $r_{\rm h}$ & $r_{\rm a}$ & $\chi^2$ & $W_0$ & $M_{\rm cl}$ & $r_{\rm h}$ & $r_{\rm a}$ & $\chi^2$\\
   & & [10$^3$M$_{\odot}$] & [pc] & [pc] & & & [10$^3$M$_{\odot}$] & [pc] & [pc] & \\
		\hline
  NGC\_188 & 4.9$\pm$0.2 & 7.3$\pm$0.3 & 6.6$\pm$0.2 & 606$\pm$358 & 28 & 4.2$\pm$0.4 & 7.4$\pm$0.3 & 6.6$\pm$0.2 & 467$\pm$339 & 31\\
  NGC\_2420 & 7.0$\pm$0.3 & 6.1$\pm$0.5 & 9.9$\pm$1.1 & 554$\pm$346 & 29 & 8.1$\pm$0.2 & 6.3$\pm$0.6 & 35.0$\pm$5.7 & 568$\pm$318 & 19\\
  NGC\_2425 & 8.2$\pm$0.2 & 4.6$\pm$0.5 & 17.8$\pm$2.2 & 204$\pm$334 & 22 & 8.1$\pm$0.2 & 6.3$\pm$0.6 & 35.0$\pm$5.7 & 568$\pm$318 & 19\\
  NGC\_2682 & 6.3$\pm$0.2 & 10.5$\pm$0.5 & 9.3$\pm$0.5 & 526$\pm$337 & 37 & 6.0$\pm$0.5 & 10.7$\pm$2.8 & 9.4$\pm$1.6 & 339$\pm$349 & 40\\
  NGC\_6791 & 5.0$\pm$0.5 & 30.7$\pm$2.0 & 12.7$\pm$1.2 & 1.9$\pm$0.4 & 47 & 5.7$\pm$0.5 & 26.8$\pm$1.4 & 10.6$\pm$0.8 & 7.9$\pm$193 & 58\\
  NGC\_6819 & 6.8$\pm$0.2 & 22.0$\pm$1.4 & 11.5$\pm$1.0 & 53.8$\pm$280 & 63 & 6.8$\pm$0.2 & 25.3$\pm$1.8 & 13.9$\pm$1.7 & 216$\pm$337 & 51\\
\hline
\end{tabular}
\end{table*}

\begin{table*}
\setlength{\tabcolsep}{1.5mm}
	\centering
	\caption{Best-fitting parameters of \limepy~models}
	\label{tab:limepymodels}
	\begin{tabular}{lccccccccc}
		\hline
		Cluster & $W_0$ & $g$ & $M$ & $r_{\rm h}$ & $r_{\rm a}$ & $r_{\rm c}$ & $r_{\rm t}$ &  $c=\log r_{\rm t}/r_{\rm c}$ & $\chi^2$\\
  & & & [10$^3$M$_{\odot}$] & [pc] & [pc] & [pc] & [pc] & & \\
		\hline
  NGC\_188 & 5.0$\pm$0.3 & 0.6$\pm$0.4 & 7.3$\pm$0.3 & 6.6$\pm$0.2 & 478$\pm$332 & 3.3 & 28.2 & 0.9 (8.7) & 27\\
  NGC\_2420 & 7.1$\pm$0.3 & 1.7$\pm$0.4 & 6.7$\pm$0.8 & 11.1$\pm$2.2 & 385$\pm$333 & 2.6 & 221.3 & 1.9 (83.8) & 28 \\
  NGC\_2425 & 8.2$\pm$1.0 & 0.6$\pm$0.5 & 4.7$\pm$1.8 & 19.4$\pm$11.7 & 485$\pm$344 & 2.3 & 126.8 & 1.7 (55.15) & 28\\
  NGC\_2682 & 6.2$\pm$0.3 & 1.1$\pm$0.5 & 10.4$\pm$0.6 & 9.3$\pm$0.5 & 371$\pm$344 & 3.2 & 73.1 & 1.3(22.6) & 36 \\
  NGC\_6791 & 4.0$\pm$0.7 & 0.9$\pm$0.3 & 29.0$\pm$2.1 & 11.8$\pm$1.3 & 1.2$\pm$0.3 & 5.2 & 269.3 & 1.7 (51.6) & 49\\
  NGC\_6819 & 6.8$\pm$0.2 & 1.9$\pm$0.2 & 24.2$\pm$1.8 & 13.0$\pm$1.4 & 498$\pm$337 & 3.3 & 416.0 & 2.1(125.8) & 51 \\
\hline
	\end{tabular}
\end{table*}

\begin{table*}
\setlength{\tabcolsep}{1.5mm}
	\centering
	\caption{Best-fitting parameters of \spes~models}
	\label{tab:spesmodels}
	\begin{tabular}{lcccccccccc}
		\hline
		Cluster & $W_0$ & $B$ & $\eta$ & $M_{\rm cl}$ & $r_{\rm h}$ & $r_{\rm c}$& $r_{\rm t}$ & fpe\tablefootmark{a} & $c=\log r_{\rm t}/r_{\rm c}$& $\chi^2$\\
  & & & & [10$^3$M$_{\odot}$] & [pc] & [pc] & [pc] & & & \\
		\hline
%  NGC\_188 & 4.1$\pm$0.4 & 1.0$\pm$0.1 & 0.8$\pm$0.1 & 8.3$\pm$0.5 & 7.3$\pm$0.5 & 3.9 & 55.0 & 0.13 & 1.1 (14.0) & 31\\
  NGC\_188 & 4.7$\pm$0.4 & 0.10$\pm$0.15 & 0.62$\pm$0.02 & 6.8$\pm$0.4 & 6.2$\pm$0.3 & 3.2 & 17.2 & 0.15 & 0.73 & 37\\
%  NGC\_2420 & 6.3$\pm$0.4 & 0.0$\pm$0.1 & 0.67$\pm$0.02 & 5.1$\pm$0.4 & 7.8$\pm$0.7 & 2.4 & 25.5 & 0.17 & 1.0 (10.4) & 23 \\
  NGC\_2420 & 6.4$\pm$0.4 & 0.0$\pm$0.02 & 0.68$\pm$0.02 & 5.1$\pm$0.4 & 7.7$\pm$0.7 & 2.4 & 25.0 & 0.18 & 1.0 & 23 \\
%  NGC\_2425 & 6.9$\pm$0.5 & 0.3$\pm$0.3 & 0.72$\pm$0.04 & 3.3$\pm$0.7 & 10.6$\pm$3.3 & 2.2 & 35.6 & 0.23 & 1.2(16.1) & 15\\
  NGC\_2425 & 7.0$\pm$0.5 & 0.02$\pm$0.09 & 0.74$\pm$0.03 & 3.0$\pm$0.5 & 9.6$\pm$2.0 & 2.1 & 30.5 & 0.23 & 1.2 & 15\\
%  NGC\_2682 & 6,0$\pm$0.3 & 0.1$\pm$0.1 & 0.64$\pm$0.02 & 9.5$\pm$0.5 & 8.3$\pm$0.5 & 3.0 & 26.6 & 0.14 & 0.9 (8.9) & 36 \\
  NGC\_2682 & 6.1$\pm$0.3 & 0.0$\pm$0.02 & 0.65$\pm$0.02 & 9.4$\pm$0.5 & 8.2$\pm$0.5 & 2.9 & 26.5 & 0.15 & 1.0 & 36 \\
%  NGC\_6791 & 4.0$\pm$0.4 & 0.8$\pm$0.2 & 0.56$\pm$0.04 & 22.2$\pm$0.2 & 8.4$\pm$0.7 & 4.5 & 30.1 & 0.14 & 0.8 (6.7) & 38\\
  NGC\_6791 & 4.7$\pm$0.4 & 0.06$\pm$0.20 & 0.65$\pm$0.02 & 20.5$\pm$0.8 & 7.9$\pm$0.3 & 4.1 & 21.1 & 0.17 & 0.7 & 38\\
%  NGC\_6819 & 6.3$\pm$0.2 & 0.0$\pm$0.1 & 0.64$\pm$0.01 & 18.9$\pm$0.4 & 9.2$\pm$0.4 & 3.1 & 30.8 & 0.15 & 1.0 (10.0) & 37 \\
  NGC\_6819 & 6.2$\pm$0.2 & 0.10$\pm$0.11 & 0.63$\pm$0.02 & 19.0$\pm$0.8 & 9.3$\pm$0.4 & 3.1 & 31.4 & 0.15 & 1.0 & 37 \\
\hline
	\end{tabular}
 \tablefoot{\tablefoottext{a}{Fraction of potential escapers.}}
\end{table*}

 The best fits for King (green dashed lines), Wilson (purple dot-dashed lines), \limepy~(blue solid lines), and \spes~(red solid lines) models are shown in Fig.~\ref{fig:model_fits}. The corresponding parameters are listed in Tables~\ref{tab:limepy_king}, \ref{tab:limepymodels}, and \ref{tab:spesmodels} for King, Wilson, \limepy, and \spes~models, respectively. Individual fits, together with their uncertainties ranges, are shown in Figs.~\ref{fig:king}, \ref{fig:wilson}, \ref{fig:limepy}, and \ref{fig:spes}, respectively. 

As expected, \limepy~models, not only in the specific King and Wilson prescriptions but also the general ones, are not able to reproduce the outskirts of the observed radial density profiles (Fig.~\ref{fig:model_fits}). The \spes~models reproduce the profiles at large radii, due to the inclusion of potential escapers, being the ones with the lowest $\chi^2$ values. The only exception is NGC\,188 for which the fit of the \limepy~model produced a slightly lower $\chi^2$ than \spes~ones. In any case, none of the used models are able to reproduce the cusp core observed in NGC\,2682 and in a less degree in NGC\,2420. On the contrary, they reproduce quite well the flat core of NGC\,6791 and the intermediate region between the core and the tidal radius for all the studied systems (Fig.~\ref{fig:model_fits}). 
  
In general, the simple King and Wilson models produce larger $\chi^2$ values than the general \limepy~models without significant differences among them except in the outskirts. The smooth variation between the King and Wilson models, and also the Woolley ones, is controlled by the variation of the truncation parameter. In our case, we found that the majority of the studied clusters are close to the King models with $g\sim$1, contrary to what was reported by \citet{deboer2019} for globular clusters, with values close to the Wilson template. The individual King models reproduce better the cluster profiles in four cases, while the Wilson ones work better for NGC\,2425 and NGC\,6819. This may be due to the fact that these systems are relatively more extended than others. As a conclusion, King models are a good first approximation to the open clusters density profiles, especially if the external regions are not included. 

A key parameter in the \limepy~models is the anisotropy parameter, $r_{\rm a}$. We found large values of $r_{\rm a}$ for all the studied clusters, which implies that the amount of anisotropy in them is quite low. The only exception is NGC\,6791 for which we found a very low value of $r_{\rm a}$. About the half mass radius, $r_{\rm h}$, which encloses half of the total mass of the system, the \spes~models values are slightly lower than those of the \limepy~ones with the only exception of NGC\,188, for which the best \limepy~model has a slightly lower value than in the best \spes~one. In both cases, the lowest value is obtained for NGC\,188. About the masses, we also found that the \limepy~models produce slightly larger values than the \spes~ones. This is explained that the \spes~models consider that a fraction of observed stars are unbound to the systems, so the amount of mass needed is lower. The obtained masses may be considered as a lower limit of the real value. Owing to the fact that we have a magnitude limit, $G\leq$18.5\,mag, we do not sample the faintest objects, which affect particularly the furthest object, NGC\,6791. Using different magnitude threshold, we check that the derived masses are comparable while there are objects below the turn-off, such as one magnitude, but they decrease when the limit is around the turn-off or above it. Moreover, we excluded binaries and blue stragglers stars from this analysis.

The core radius, $r_{\rm c}$, and truncation or tidal, $r_{\rm t}$ radius are not obtained from the fits, but they are computed as a function of the best results for both \limepy~and \spes~models, respectively. Derived core radii for both families of models are similar, showing the same trend that the \limepy~models values are slightly larger than the \spes~models ones, again except for NGC\,188. NGC\,6791 shows the largest $r_{\rm c}$, while NGC\,2425 has the smallest one. Tidal radii show a similar tendency. However, the values derived for \limepy~models are very different among clusters, from the 28.8\,pc of NGC\,188 to the 416\,pc of NGC\,6819. Only NGC\,188 and NGC\,2682 show a flattening in the outermost radii studied, most probably due to the algorithm not being able to properly place the end of the cluster. On contrary, the \spes~models provided smaller tidal radii, between 25\,pc for NGC\,2420 and 55\,pc for NGC\,188. This implies that the objects located outside these radii are extra-tidal stars which are probably escaping from the clusters. In fact, according to \spes~models between 13\%, for NGC\,188, and 23\%, for NGC\,2425, of the observed stars in each system may have enough energy to escape from them.

Other works have studied the radial density profiles of the clusters in our sample, mainly by fitting them with the analytical King profile model. The majority of them have been performed in recent years, taken advantage of the different \gaia~ data releases \citep{Gao2018,gao2022_ngc188,zhong2022,angelo2023,cordoni2023}. Together with the variety of algorithms used to calculate the membership probabilities, our main difference is the studied area around the cluster centre, significantly larger than in the majority of the other cases. In general, the values for the core radius reported by these works are of the order, or slightly smaller than, the values found here. There are larger discrepancies in the tidal radius determination. This is explained in part by the small area, relative to us, covered by several of these works, and by the presence of unbound stars in the same way as the differences found above between \limepy~and \spes~models. To our knowledge, the recent work by \citet{angelo2023} is the only one that derived masses for the studied clusters, four in common with us: NGC\,188, NGC\,2682, NGC\,6791, and NGC\,6919. Using a different approach, the masses derived for these clusters are in good agreement, within the uncertainties, with the ones obtained here from the \limepy~models.

\section{Mass segregation}\label{sec:mass_segregation}

It has been widely reported in the literature that old open clusters are segregated in mass \citep[e.g.,][]{mathieu1984_m11}, with massive objects concentrated in the central region while low mass stars are dispersed in the outskirts. In spite of some evidence of primordial mass segregation in very young massive clusters \citep[e.g.][]{kim2006,stolte2006}, in old open clusters this should be a direct consequence of their internal dynamics.

\begin{figure*}
	\includegraphics[width=\textwidth]{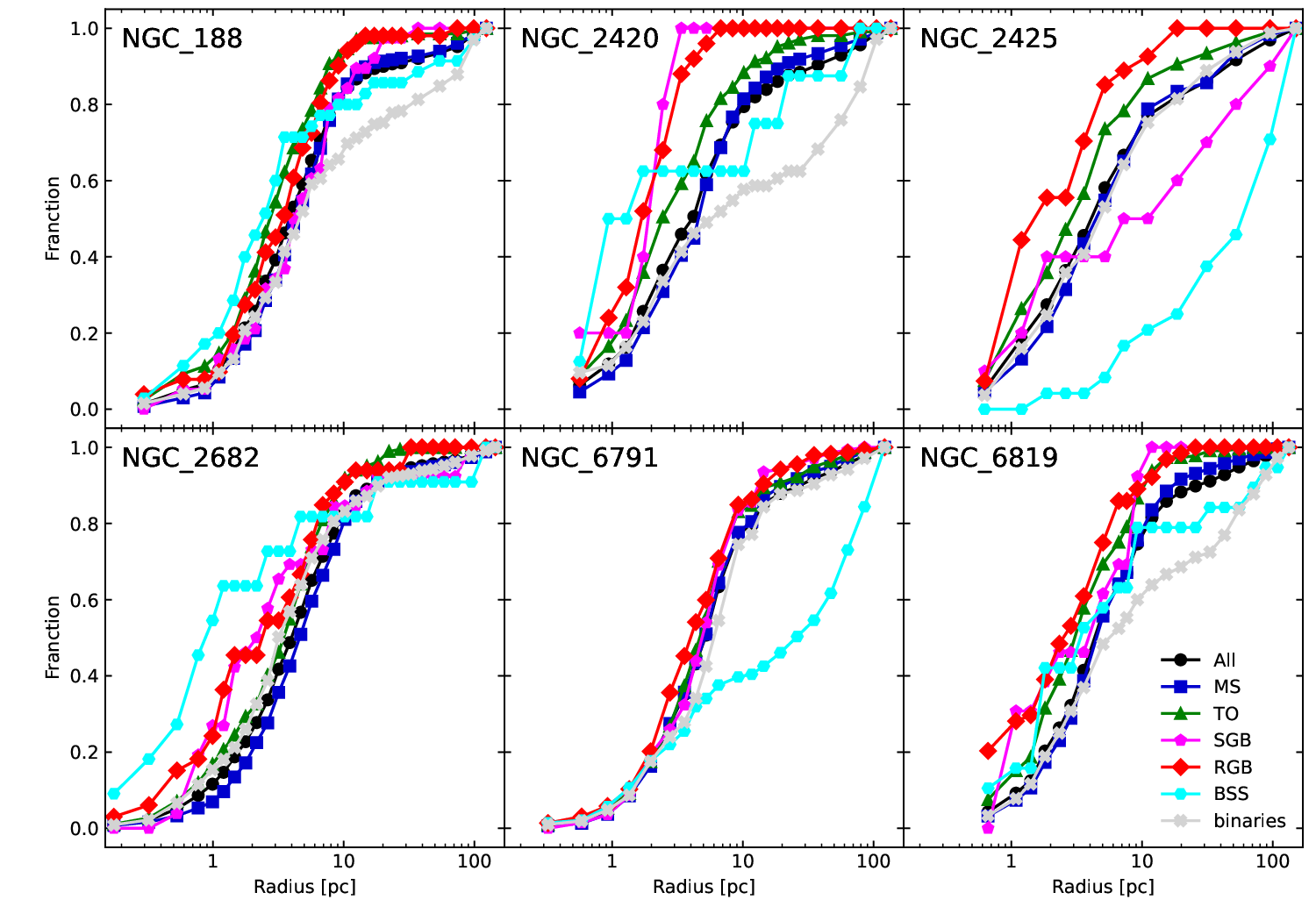}
    \caption{Cumulative projected radial distribution of the different populations identified in each of the clusters, normalised to the total number of stars in each population. Error bars have not been plotted for clarity.}
    \label{fig:CDs}
\end{figure*}

 \begin{table*}
\setlength{\tabcolsep}{0.5mm}
	\centering
	\caption{Summary of cumulative projected radial distribution of different populations.}
	\label{tab:percentiles}
	\begin{tabular}{lcccccccccccccccccccccc}
		\hline
		Population & \multicolumn{3}{c}{All} & \multicolumn{3}{c}{MS} & \multicolumn{3}{c}{TO} & \multicolumn{3}{c}{SGB}  & \multicolumn{3}{c}{RGB}  & \multicolumn{3}{c}{BSS}& \multicolumn{3}{c}{Binaries} &  $\Lambda_{\rm MSR}$(15)\\
		 & \textit{N} & $r_{15\%}$ & $r_{85\%}$ & \textit{N} & $r_{15\%}$ & $r_{85\%}$ & \textit{N} & $r_{15\%}$ & $r_{85\%}$ & \textit{N} & $r_{15\%}$ & $r_{85\%}$ & \textit{N} & $r_{15\%}$ & $r_{85\%}$ & \textit{N} & $r_{15\%}$ & $r_{85\%}$ & \textit{N} & $r_{15\%}$ & $r_{85\%}$& \\
  & & [pc] & [pc] & & [pc] & [pc] & & [pc] & [pc] & & [pc] & [pc] & & [pc] & [pc]& & [pc] & [pc]& & [pc] & [pc]& \\
		\hline
NGC\_188   & 1061 & 1.6 & 12.0 & 533 & 1.7 & 11.3 &204 & 1.3 & 7.3 &38 & 1.5 & 11.0 &51 & 1.5 & 7.8 &35 & 0.9 & 16.0 &200 & 1.7 & 69.6 & 1.6$\pm$0.9\\
NGC\_2420  &  681 & 1.2 & 17.2 & 427 & 1.5 & 13.8 &105 & 1.0 & 9.2 &5 & 1.2 & 2.9 &25 & 1.0 & 3.7 &8 & 1.0 & 19.8 & 111 & 1.1 & 90.7 & 5.5$\pm$0.6\\
NGC\_2425 &  425 & 1.4 & 35.4 & 175 & 1.7 & 27.9 &108 & 1.1 & 12.1 &10 & 1.4 & 68.4 &27 & 1.0 & 5.6 &24 & 9.3 & 144.0 &81 & 1.5 & 32.7 & 3.7$\pm$0.6\\
NGC\_2682 & 1158 & 1.4 & 12.0 & 676 & 1.7 & 13.1 &185 & 1.0 & 8.7 &26 & 0.9 & 10.2 &33 & 0.7 & 7.7 &11 & 0.5 & 11.7 &227 & 1.0 & 12.5 & 2.1$\pm$0.7\\
NGC\_6791  & 2481 & 2.1 & 16.2 &1001 & 2.2 & 14.5 &492 & 2.1 & 12.4 &139 & 2.0 & 11.3 &292 & 2.1 & 11.2 &141 & 2.0 & 99.1 &416 & 2.1 & 17.5& 1.3$\pm$0.7\\
NGC\_6819  & 2021 & 1.7 & 16.8 &1325 & 1.8 & 14.2 &229 & 1.1 & 10.0 & 13 & 1.0& 8.5 &65 & 0.8 & 7.0 &20 & 1.1 & 40.1 & 369 & 1.7 & 74.8& 2.2$\pm$0.6\\
\hline
\end{tabular}
% \tablefoot{Values obtained from \citet{cantatgaudin2020} derived from \gaia DR2, except for proper motions and parallaxes, which have been recomputed using \gaia DR3. \tablefoottext{a}{The projection on sky corresponds to a physical radius of 150\,pc for each system.}}
\end{table*}

In order to investigate the mass segregation of the studied clusters, we have obtained the cumulative projected radial distribution of the different stellar populations normalised to the total number of the objects in each of them. In spite of the uncertainties in our membership determinations for blue stragglers and binaries, we included both populations in our analysis \citep[see][for details]{m67paper}. According to Fig.~\ref{fig:CDs}, all the studied clusters are segregated in mass, with the red giant branch more concentrated than the less massive objects in the main sequence. The only exception is the central part of NGC\,6791 since within the innermost 2\,pc there is no clear separation among the different stellar populations. Mass segregation has already been reported in the literature for the majority of the studied clusters, mostly based on membership determination from pre-\gaia era: NGC\,188 \citep[e.g.][]{Bonatto05, Geller2008}, NGC\,2420 \citep[e.g.][]{Leonard1988, Peikov2002, Paparo}, NGC\,2682 \citep[e.g.][]{Lbalaguer2007, Geller2015,Gao2018}, NGC\,6791 \citep[e.g.][]{Gao2020, Platais2011}, and NGC\,6819 \citep[e.g.][]{Kalirai2001, Kang2002, Yang2013}. In general, the global distributions are mainly due to the main sequence stars, whereas turn-off and subgiant branch objects are more concentrated. To quantify this statement, we have computed the radii, which contain the 15\% and 85\% of each population (Table~\ref{tab:percentiles}). The concentration of the different populations changes from one cluster to the other. For example, in the case of NGC\,2682, the more massive red giants and subgiants stars are notoriously more concentrated towards the innermost parts than the other populations. It seems that in the case of NGC\,188, the turn-off objects are more concentrated than red giant and subgiants objects. This could be explained by the fact that for NGC\,188 the mass difference between red giant and main sequence turn-off objects should be lower than in the case of NGC\,2682 since it is about 4\,Ga older. For NGC\,2420, there is a non-negligible quantity of binaries in the outermost radii. Finally, the subgiant population in NGC\,2425 exhibits an unexpected behaviour, flattening between $\sim$1 and 10\,pc, but it could be and effect due to being based on a few objects.

Particularly interesting is the case of the blue stragglers, which are typically more concentrated than even the red giants. The mechanisms for the formation of blue stragglers are still not fully understood. These stars are thought to derive from normal main sequence stars that have increased in mass above a single star mass typical of the turn-off through mass transfer, mergers \citep[e.g.][]{mccrea1964,paczynski1971}, or collisions in binary systems \citep[e.g.][]{hilss1976}. As reported by \citet{m67paper}, these objects show a bimodal distribution in the majority of the studied clusters, as observed in globular clusters \citep[e.g.][]{ferraro1997,lanzoni2007} and predicted by dynamical simulations \citep[e.g.][]{mapelli2004}. 

In order to quantify and compare the degree of mass segregation among the studied clusters, we used the method proposed by \citet{allison2009}. It consists of comparing the lengths of the minimum spanning tree (MST) of the most massive stars of a cluster and a set of the same number of randomly chosen objects between all populations. An MST of a set of points is the path connecting all the points, with the shortest possible path length and without any closed loops. In a given set of points, only one MST can be drawn. In our case, we defined the mass segregation ratio (MSR) as:

\begin{equation*}
    \Lambda_{\rm MSR} (N)=\frac{\langle l_{\rm random} \rangle}{{\langle l_{\rm massive} \rangle}}
\end{equation*}
\noindent
\noindent where $\langle l_{\rm random} \rangle$ and $\langle l_{\rm massive} \rangle$ are the average lengths of the MST of N randomly selected stars between all and massive samples, respectively. The average lengths $\langle l_{\rm random, massive} \rangle$ are calculated over 100 iterations, where at each iteration, we drew a different subsample of random stars allowing us to simultaneously calculate $\sigma_{\rm random, massive}$, the standard deviations of the lengths of the MST. We computed the uncertainty of $\Lambda_{\rm MSR}$ as the square root of the quadratic sum of $\sigma_{\rm random}$ and $\sigma_{\rm massive}$.

We computed the MST by using the \texttt{csgraph} routine implemented in the \texttt{scipy} \python~ module \citep{scipy}. We consider the red giant sample as representative of the massive stellar population in each cluster. After several trials, we assume $N=15$ stars. In principle, a $\Lambda_{\rm MSR}$ greater than one means that the massive population is more concentrated than a random sample, and therefore that the cluster shows signs of mass segregation. The obtained values are listed in the last column of Table~\ref{tab:percentiles}. In all the studied clusters, the red giants are more concentrated than random selected stars. Noticeable are the larger values found for NGC\,2420 and NGC\,2425, which are the less massive clusters in our sample (see Table~\ref{tab:spesmodels}) and also among the youngest, where a larger mass difference between giants and unevolved stars is expected. On the other hand, the most massive systems have lower $\Lambda_{\rm MSR}$ values. In fact, the lowest value is found for the most massive system in our sample, NGC\,6791,  according to the previous section results, although this is also the furthest cluster which implies that our sample only includes members in the upper MS, and therefore, we are not sampling as many low-mass stars as in other clusters. 

\section{Physical parameters of the clusters}\label{sec:analysis}

\subsection{Half-mass relaxation time}

The dynamical evolution inside the clusters is related to the encounters among stars. After losing a certain amount of members, the relaxation time is the moment in which the stellar system reaches the equilibrium. In other words, this is also defined as the time required for a star to lose the memory of its initial conditions. The relaxation time is locally defined, and it can vary by several orders of magnitude in different regions of a single cluster. For this reason, the relaxation time is widely calculated in reference to the system's half-mass radius, the so-called half-mass relaxation time, $t_{\rm rh}$. 

The half-mass relaxation time can be derived according to \citet{mclaughlin2005} as:

\begin{equation*}
    t_{\rm rh}[a]=\left[\frac{2.06\times10^6}{\ln{\frac{0.4 M_{cl}}{m_*}}}\right]m_*^{-1} M_{cl}^{1/2}r_{h}^{3/2}
\end{equation*}

\noindent where $M_{\rm cl}$ and $r_{\rm h}$ are expressed in M$_{\odot}$ and parsecs, respectively. In this section, we consider only the values derived from the \spes~models since they reproduce better the radial density profiles in comparison to the \limepy~models. The average current stellar mass in the cluster, $m_*$, is defined as $M_{\rm cl}/N_{\rm cl}$ being $N_{\rm cl}$ the number of stars in the cluster. Assuming the \citet{kroupa2001} initial mass function, $m_*$ would have a value of 0.54\,M$_{\odot}$. Considering that the most massive stars have already died, we assume $m_*\sim$0.5$\pm$0.1\,M$_{\odot}$ to allow some deviations. Obtained values are listed in Table~\ref{tab:othercalcs} together with their dynamical ages $\tau_{\rm rh}=\log age/t_{\rm rh}$.

Recently, \citet{angelo2023} have determined half-mass relaxation time for a sample which includes four systems in common with us. Except for NGC\,6791, our values are larger than those derived by \citet{angelo2023} for the four clusters in common, mainly motivated by the fact that we are using larger input clusters' masses and half-mass radii since our study covers a wider area and contains a larger number of members.

\subsection{Jacobi radius}

Stellar clusters do not live in isolation, but they are orbiting the Galaxy inside its tidal field. The Jacoby radius, $R_{\rm J}$, also called Hill or Roche radius, of a stellar cluster is the maximum distance inside which a star is still bound gravitationally to the system taking into account the external Galactic tidal potential. Therefore, it can be considered as a prediction of the tidal radius. According to \citet{King62}, $R_{\rm J}$ can be derived as:

\begin{equation*}
    R_{\rm J}=R_{\rm p}\left(\frac{M_{\rm cl}}{(3+e)M_{\rm g}}\right)^{1/3}
\end{equation*}

\noindent where $R_{\rm p}$ is the perigalactic radius defined as $R_{\rm p}=R_{\rm apo}(1-e)$ with $R_{\rm apo}$ and $e$ being the apocentre radius and the eccentricity of the cluster orbit, and $M_{\rm cl}$ and $M_{\rm g}$ the masses of the cluster and the Galaxy enclosed within the orbit of each cluster, respectively.

We use the orbits parameters determined by \citet{carrera2022occaso} except for NGC\,2425 for which we assume the values determined by \citet{tarricq2021}. In both cases, the clusters' orbits were computed using the \texttt{galpy} package \citep{galpy}. In order to determine the mass enclosed, $M_{\rm g}$, within the orbit of each system, which is model dependent, we also take advantage of the \texttt{galpy} package assuming the \textit{MW2014} potential for the Milky Way \citep[see][for details]{galpy}.

The derived Jacobi radius are listed in Table~\ref{tab:othercalcs}. Using other potentials available in the literature, such as that of \citet{mcmillan2017}, yields similar values. Increasing the mass of the clusters, for example using the values derived from \limepy~models, yields slightly larger Jacobi radius. On the other hand, an increase of the Galaxy mass enclosed inside the cluster orbit tends to reduce $R_{\rm J}$. 

Alternatively, the Jacobi radius can be derived from the tidal tensor of the total potential \citep{renaud2011}.  In general, their values are lower than those derived here, except for NGC\,6791. This discrepancy may be due to the fact that our masses, from \spes~models, are larger than those derived by \citet{angelo2023}. In any case, neither our $R_{\rm J}$ estimations nor the ones by \citet{angelo2023} are as large as the tidal radius determined by both \spes~ and \limepy~models.

%  derived values are significantly different than those derived here, being between twice a three times larger: NGC\,188 24.6$\pm$1.8\,pc; NGC\,2682 17.8$\pm$2.0\,pc; NGC\,6791 38.1$\pm$13.2\,pc; NGC\,6819 23.2$\pm$2.1\,pc
\begin{table*}
\setlength{\tabcolsep}{0.6mm}
	\centering
	\caption{Physical properties obtained using clusters' masses derived from \spes~models.}
	\label{tab:othercalcs}
	\begin{tabular}{lcccccc}
		\hline
		Cluster & $t_{rh}$ & $\tau_{rh}$ & $R_{\rm J}$ & $M_{\rm ini}$ & $t_{\rm dis}$ & $t_{\rm disc}$\\
   & [Ga] & [dex] & [pc] & [10$^3$\, M$_{\odot}$] & [Ga] & [Ma]\\ 
		\hline
  NGC\_188 & 0.7$\pm$0.1 & 1.01$\pm$0.08 & 32.1$\pm$1.1 & 26.1$\pm$2.2 & 13.4$\pm$0.8 &44$\pm$1\\
  NGC\_2420 & 0.7$\pm$0.1 & 0.42$\pm$0.10 & 32.4$\pm$1.3 & 9.7$\pm$0.7 &  7.9$\pm$0.5 & 30$\pm$2\\
  NGC\_2425 & 0-9$\pm$0.2 & 0.35$\pm$0.10 & 27.7$\pm$2.4 & 9.5$\pm$1.8 & 4.0$\pm$0.6 & 54$\pm$3\\
  NGC\_2682 & 1.0$\pm$0.1 & 0.55$\pm$0.08 & 35.0$\pm$0.9 & 23.8$\pm$2.1 &  9.5$\pm$1.0 & 38$\pm$2\\
  NGC\_6791 & 1.4$\pm$0.2 & 0.78$\pm$0.08 & 41.0$\pm$1.9 & 70.4$\pm$4.5 & 16.3$\pm$0.8 & 26$\pm$2\\
  NGC\_6819 & 1.5$\pm$0.2 & 0.11$\pm$0.07 & 40.8$\pm$1.5 & 34.8$\pm$1.9 &  9.6$\pm$1.4 & 12$\pm$1\\
\hline
	\end{tabular}
\end{table*}

\subsection{Initial mass}

A stellar cluster losses mass due mainly to stellar evolution and tidal disruption. \citet{lamers2005} showed that the fraction of the cluster initial mass, $M_{\rm ini}$, lost due to stellar evolution, ($\Delta M)_{\rm ev}$, in the \texttt{GALEV} \citep[Galaxy Evolutionary Synthesis,][]{kotulla2009} models can be written as $\frac{(\Delta M)_{\rm ev}}{M_{\rm ini}}\equiv q_{\rm ev}$, which can be approximated by:

\begin{equation*}
\log q_{ev}(t)=\left(\log age -a_{ev} \right)^{(b_{ev})}+c_{ev} \textrm{\;for\;} t>12.5\textrm{\,Ma}
\end{equation*}

\noindent where $a_{\rm ev}$, $b_{\rm ev}$, and $c_{\rm ev}$ are coefficients which slightly depend on metallicity according to Table~1 by \citet{lamers2005}. Metallicities are obtained from iron abundances using the values provided by Carbajo-Hijarrubia et al. {\sl in prep.} except for NGC\,2425 which was obtained from \citet{randich2022ges_oc}.

More generally, the evolution with time of the mass of a cluster which has survived the infant mortality, age $\geq$10$^7$\,a, can be described as:

\begin{equation*}
    \frac{d M_{\rm cl}}{d t}=\left(\frac{d M_{\rm cl}}{d t}\right)_{ev}+\left(\frac{d M_{\rm cl}}{d t}\right)_{dyn}
\end{equation*}

\noindent where the first term describes mass loss due to stellar evolution and the second by disruption.

The mass decrease of a cluster can be approximated very accurately by:

\begin{equation}\label{eq:mu}
\mu (t; M_{\rm ini}) \equiv \frac{M_{\rm cl}}{M_{\rm ini}} \sim \left\{ \left[ \mu_{\rm ev}(t) \right]^{\gamma} - \frac{\gamma}{M^{\gamma}_{\rm ini}} \frac{t}{t_0}\right\}^{1/\gamma}   
\end{equation}

\noindent where masses are expressed in M$_{\odot}$, $\mu_{\rm ev}(t)=1-q_{\rm ev}(t)$, and $\gamma$=0.62 according to \citet{lamers2005}. The constant $t_0$ depends on the galaxy potential in which the cluster is moving on, and on the eccentricity, $e$, of its orbit. From \citet{lamers2005b}, it can be derived from: 

\begin{equation*}
t_0=C_{\rm env,0}(1-e) 10^{-4 \gamma} \rho_{\rm amb}^{-0.5}
\end{equation*}

\noindent where $C_{\rm env,0}$=810\,Ma for the Milky Way \citep[see][for details]{lamers2005b}. $\rho_{\rm amb}$ is the ambient evaluated at the position of the cluster, expressed in M$_{\odot}$\,pc$^{-3}$. It was determined for each cluster with the \texttt{galpy} package, assuming the \textit{MW2014} potential \citep{bovy2016}.

The initial cluster mass can be easily derived by manipulating Eq.~\ref{eq:mu} as:

\begin{equation*}
M_{\rm ini} \simeq \frac{1}{\mu_{ev}}\left\{ M_{\rm cl}^{\gamma}+\gamma \frac{t}{t_0} \right\}^{1/\gamma}     
\end{equation*}

The obtained $M_{\rm ini}$ values are listed in Table~\ref{tab:othercalcs}. In comparison with \citet{angelo2023}, who determined initial masses following a similar procedure, our estimations are slightly larger, except for NGC\,6791, the most massive system in our sample, which is significantly smaller. Together with the different input clusters' masses, the only difference between both studies is the treatment of the Galaxy tidal field.    

Finally, \citet{baumgardt2003} found that the disruption time of a stellar cluster can be expressed as a function of $M_{\rm ini}$ as $t_{\rm dis}=t_0 M_{\rm ini}^{\gamma}$ with $t_{\rm dis}$ in years. Derived values are also listed in Table~\ref{tab:othercalcs}. Obtained values are discussed in the following section.

\section{Physical interpretation}\label{sec:discussion}

%\begin{figure}[h!]
%   \centering
   %\includegraphics[width=\columnwidth]{fig_relations_mass.eps}
%   \caption{Top: fraction of potential escapers as a function of the ratio of the current clusters' mass relative to the initial masses, colour coded as a function of age. Bottom: fraction of potential escapers as a function of the clusters' masses, colour coded by their current distances to the Galactic plane. }
%   \label{fig:relations_mass}
%\end{figure}

\subsection{Dynamical evolutionary stages}

\begin{figure*}[h!]
   \centering
   \includegraphics[width=\textwidth]{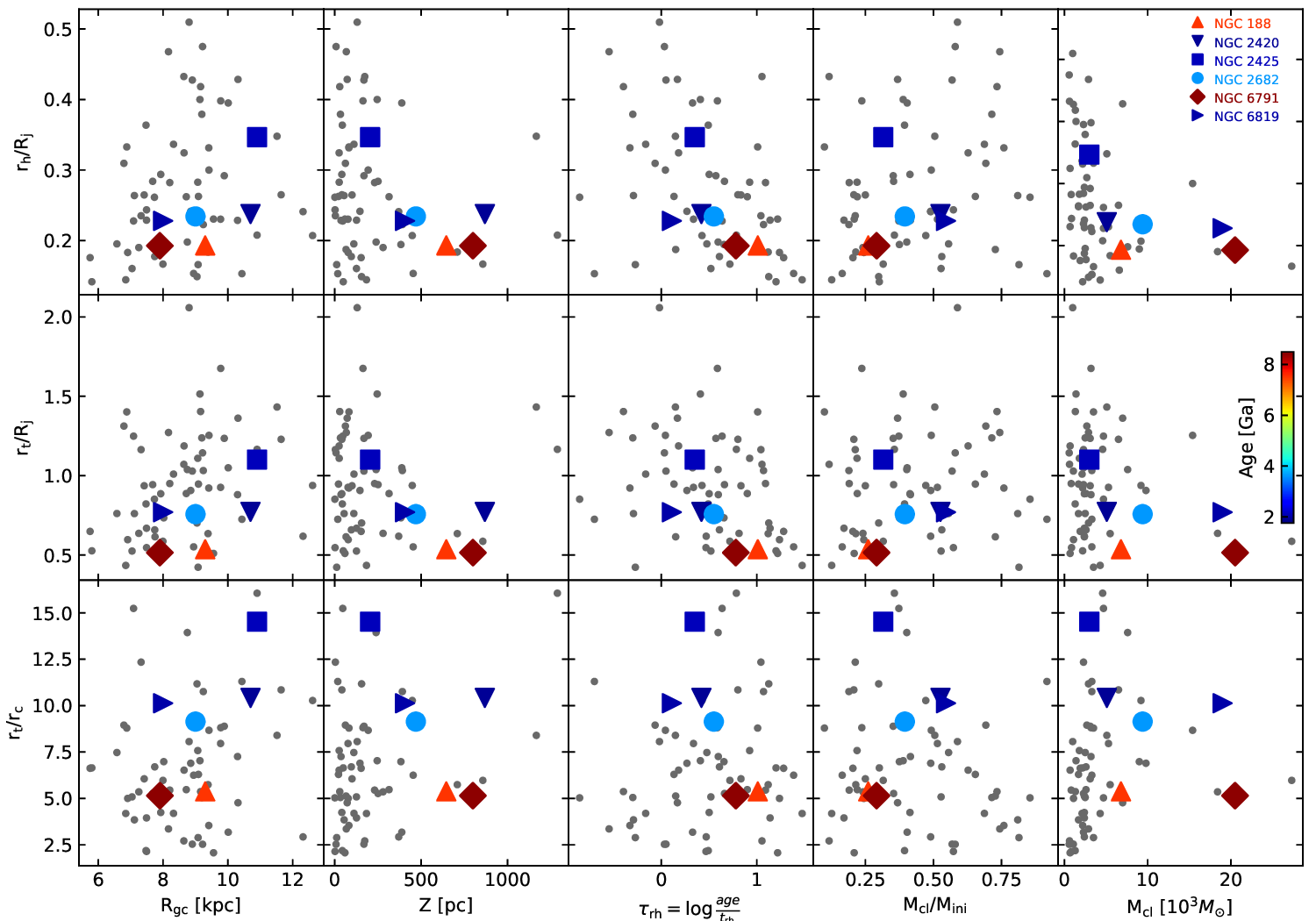}
   \caption{r$_{\rm h}$/R$_{\rm J}$ (top), r$_{\rm t}$/R$_{\rm J}$ (middle), and r$_{\rm t}$/r$_{\rm c}$ (bottom) as a function of R$_{\rm gc}$, Z, $\tau_{\rm rh}$, M$_{\rm cl}$/M$_{\rm ini}$, and cluster masses. Clusters studied here are labelled with different colours, as a function of ages, and symbols to facilitate their identification. Grey dots are the clusters studied by \citet{angelo2023}.}
   \label{fig:relations_radii}
\end{figure*}

We discuss the dynamical stages of the studied clusters as a function of the derived structural parameters (r$_{\rm h}$, r$_{\rm t}$, r$_{\rm c}$), relaxation times ($\tau_{\rm rh}$), Jacobi radii (R$_{\rm J}$), location in the disc (R$_{\rm gc}$, Z), and evolution-related parameters (ages, stellar masses). Our results are summarised in the different panels of Fig.~\ref{fig:relations_radii}. In order to compare to the global trends described by open clusters, we have also plotted the results obtained by \citet{angelo2023}. These authors studied four of the clusters in our sample: NGC\,188, NGC\,2682, NGC\,6791, and NGC\,6819. We recovered slightly large structural parameters, mainly due to the fact that we are covering a large area around the clusters. In either case, the results obtained for these systems, mainly the ratios among the different radii, are compatible within the uncertainties.

The two oldest clusters in our sample, NGC\,188 and NGC\,6791, are also the most dynamically evolved systems according to their dynamical ages, $\tau_{\rm rh}>$0.7\,dex. These systems have galactocentric distances between 8 and 9\,kpc from the centre of the Galaxy. They are located more than 500\,pc above the Galactic plane, and therefore, less affected by the destructive influence of the disc. This is in agreement with the idea that cluster of similar ages are preferentially located outside the dense disc, which may plays a role in their survival \citep[e.g.][]{friel2013,cantatgaudin2020}. These two systems also have the lower r$_{\rm t}$/r$_{\rm c}$, which can be explained by the fact that they have reduced their sizes throughout their lives by losing a significant part of their outskirts. The remnants that we observe nowadays would be mainly their cores. This hypothesis is supported by the fact that they have lost more than 70\% of their birth masses. Moreover, these two systems show the lower mass segregation ratio, $\Lambda_{\rm MSR}$. This can be a consequence of the fact that they have lost a significant fraction of their less massive stellar components. They also show the lowest r$_{\rm h}$/R$_{\rm J}$ and r$_{\rm t}$/R$_{\rm J}$ ratios, implying that they are well inside their Jacoby radius. This would mean that their evolution is mainly dominated by their internal dynamics and not by the influence of the Galactic tidal potential. This result is in agreement with the N-body simulation's prediction that lobe Roche underfilling clusters will survive a number of relaxation times \citep[e.g.][]{gieles2008}.

NGC\,6819 is a 1.9\,Ga old and very massive system, M$_{\rm cl}$=19$\times10^3$\,M$_{\odot}$, located at 8\,pc from the Galactic centre with has lost less than 50\% of its initial mass. NGC\,2682 is older, 3.6\,Ga, but less massive located at a similar height above the Galactic plane but about a kpc further, which have lost a somewhat larger fraction of its birth mass, $\sim$60\%. NGC\,2682 is slightly more concentrated, which may means that this cluster is more dynamically evolved, as evidenced by its dynamical age. In spite of their significantly different initial and nowadays masses, both systems have similar dissolution times due to the fact that the most massive one is also located at an innermost galactocentric distance.

On the other hand, NGC\,2420, which almost the same age as NGC\,6819, is located much further, R$_{\rm gc}$=10.7\,kpc, and also at a larger distance from the plane, Z=869\,pc, than NGC\,2682 and NGC\,6819. This cluster is twice and four times less massive than NGC\,2682 and NGC\,6819, respectively. However, its dissolution time is only a little lower mainly due to its location. These three clusters have similar $r_{\rm t}/R_{\rm J}$ ratios of about 0.8, which means that they occupy almost all their Roche volumes, and therefore, their dynamical evolutions are modulated by the Galactic tidal field.

Finally, NGC\,2425 is the furthest cluster in our sample but also the closest to the Galactic plane. It has an age similar to those of NGC\,2420 and NGC\,6819. This cluster shows clear hints of dissolution. On one hand, it has a $r_{\rm t}/R_{\rm J}$ ratio larger than one, meaning that it completely fills its Roche volume since it may have been more exposed to external tidal forces \citep[e.g.][]{ernst2015}. It is also the system with the lowest concentration, which may mean that the stars that form this system are less gravitationally bound. In fact, we found a significant larger fraction of potential escapers for NGC\,2425 than for the other clusters in our sample. Though, NGC\,2425 is coeval, was formed with a similar initial masses, and it is located at similar galactocentric distance than NGC\,2420, it has lost a significant larger fraction of its mass motivated by their different distances to the galactic plane. Taken into account its age, NGC\,2425 will dissolve in only about 2.5\,Ga from now.

Although our work is based in a small sample of only six clusters, we can obtain valuable conclusions which should be confirmed from larger samples. Our results reinforce the idea that the initial mass and the location in the Galaxy, mainly above or below the disc, play a fundamental role in the survival of open clusters. Those systems that do not fill completely their Roche volumes may survive longer, either they were born more concentrated or they lost a significant fraction of their outskirts. In fact, our results suggest that the concentration of the cluster increase with age.

\subsection{Disc crossing times}

In Sect.~\ref{sec:density_profiles} we reported the existence of tails in several of the studied clusters, such as NGC\,188. In order to investigate the influence of the Galactic plane in the morphology of the studied systems, we have estimated the last disc crossing. For this purpose, we use the orbits determined by \citet{carrera2022occaso} for all clusters except NGC\,2425, for which we use \citet{tarricq2021}. In both cases, the orbits were derived using the \python~\texttt{galpy} package \citep{galpy} and the \textit{ MW2014} potential. We refer the reader to the original papers for the details. We do not attempt to derive new orbits since the available values are the same, within the uncertainties, to those used in the orbit determination.

 The times of the last disc crossing of each cluster are listed in the last column of Table~\ref{tab:othercalcs}. NGC\,6819 has crossed the disc very recently, only about 12\,Ma ago, which may be the explanation of its elongation. Moreover, it has relatively large disc pass frequency since it crosses the disc about every 50\,Ma. Among the other clusters, NGC\,6791 has passed the disc about 26\,Ma ago, but there is no sign of elongation. The rest of the clusters have crossed the plane more than 30\,Ma ago. In fact, the larger value is obtained for NGC\,2425 but since this cluster do not reach a significant distance to the plane, its orbit is embedded inside the disc, and therefore, the last disc crossing is not relevant.

\begin{figure}[h!]
   \centering
   \includegraphics[width=\columnwidth]{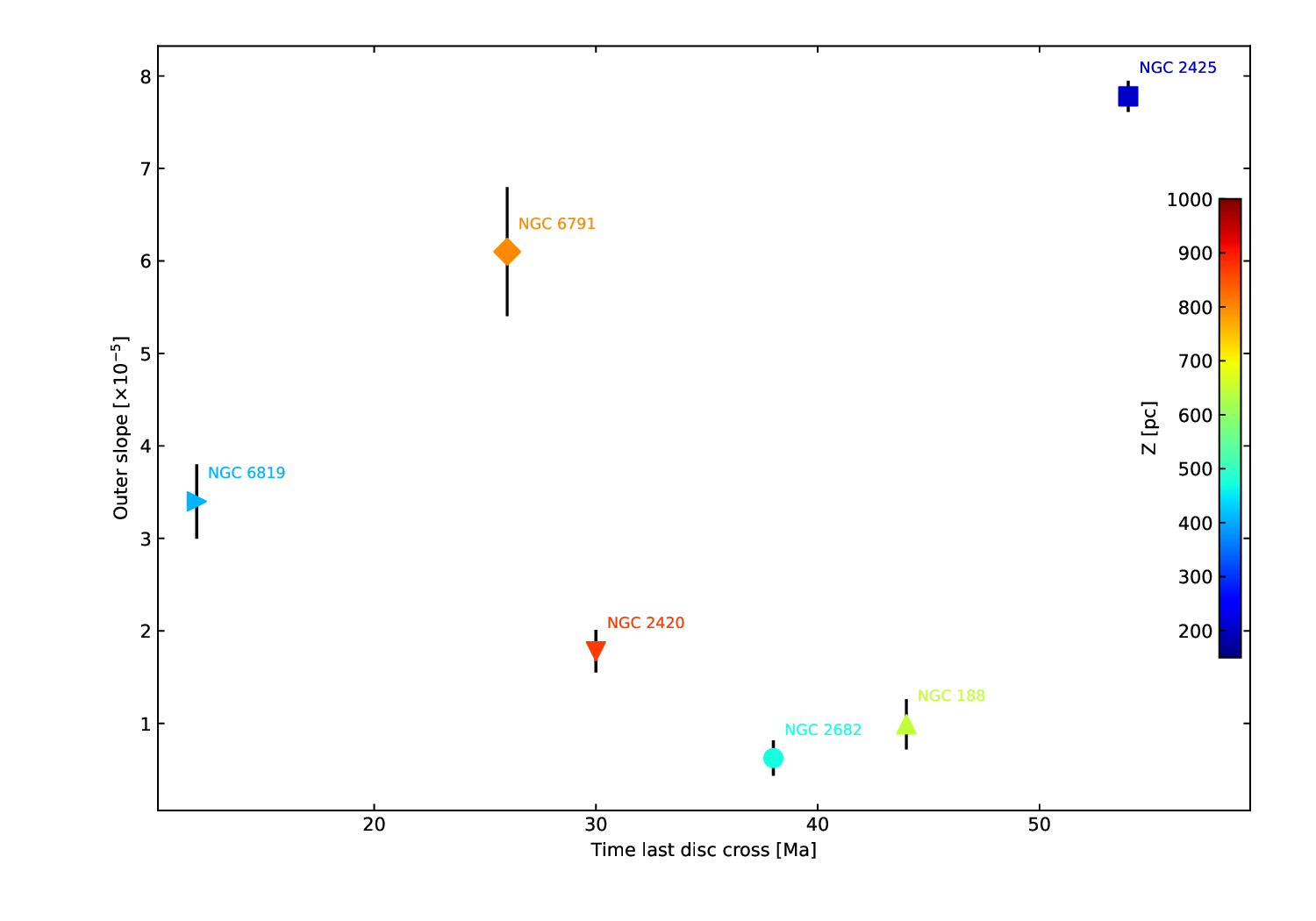}
   \caption{The slope of the outermost region of the radial density profile as a function of the last time the clusters crossed the disc. The clusters are colour coded as a function of their actual distance to the Galactic plane, using the same symbols as in previous figures.}
   \label{fig:disc_cross_slope}
\end{figure}

 In order to evaluate this impact, we have computed the slopes of the radial density profiles of the studied clusters simply using a linear least square fitting. The results are shown in Fig.~\ref{fig:disc_cross_slope}. Although the uncertainties in these external regions are large due to the small number of stars sampled, there is a hint of a correlation between the slope of the external region and the time that makes the clusters cross the disk for the last time without any clear relation with the actual distance to the Galactic plane. However, due to the small number of systems in our sample, this result should be confirmed by larger samples.

\subsection{A cusp-core dichotomy for open clusters?}

There are also clear differences in radial density profiles of the central regions among the studied clusters. In principle, these central regions are less affected by the external perturbations, and therefore, they may be dominated by the gravitational encounters of their members \citep[e.g.][]{oleary2014}. A direct consequence of two-body relaxation is the redistribution of the stars, the mass segregation, and modification of their velocities due to the exchange of energy and angular momentum.
In fact, the ages of the studied clusters are at least twice their $t_{\rm rh}$ and therefore, the two-body relaxation has had enough time to act. All the clusters show clear signs of mass segregation, particularly in their cores, with mass segregation rations, $\Lambda_{\rm MSR}$, larger than two. The exception are NGC\,188 and NGC\,6791, with the lowest mass segregation ratios, which, as discussed above, should be related to the fact that these systems have lost a significant fraction of their less massive members. 

Interestingly, the clusters with the lower $\Lambda_{\rm MSR}$ value show flat profiles in their central regions, particularly clear in the case of NGC\,6791. Meanwhile, for systems with higher values a power-law increase is apparent, the so-called cusp profile, which is specially clear in the case of NGC\,2682. From our limited sample, the systems with a larger dynamical age show a core morphology.

The cusp-core dichotomy in the central density profiles is well-known among the globular clusters, more massive and older than the open ones \citep[e.g.][]{mclaughlin2005}. Although there are still some caveats about the mechanisms responsible for the cuspy profiles, it is widely assumed that clusters with such profile have had complex internal dynamics \citep[e.g.][]{meylan1997}. They are associated with systems that have suffered a gravothermal collapse, in which the increase of the central density becomes dramatic, the so-called core-collapse \citep{lyndenbell1968}. However, the timescale for this event to happen is lower than in the case of globular clusters, so the oldest open systems could have experienced it \citep{spitzer1987}. 

For globular clusters, several features have been related to cusp systems in comparison with core ones, such as a higher central density, a large binary fraction, a significant blue straggler population, and a high number of X-ray sources \citep[e.g.][]{meylan1997}. Binary fraction between 20\% and 40\% have been reported in the literature for the clusters in our sample \citep[e.g.][]{bedin2008,milliman2014,thompson2021} but it becomes at 70\%$\pm$17\% for the central region of NGC\,2682 \citep{geller2021} being 32\%$\pm$3\% for NGC\,6791 \citep{bedin2008}. However, \citet{rain2021} reported the largest blue straggler populations for NGC\,6791 and NGC\,188 with 48 and 22 objects, respectively. NGC\,2420, NGC\,2682 and NGC\,6819 contain 3, 11, and 15 blue straggler stars, respectively \citep{rain2021}. These numbers are in agreement with our results, in spite our procedure is not the more adequate for these stars. From X-ray observations, it has been reported in the literature that NGC\,2682 contains between 2 and 7 times more active binaries, normalised to the cluster mass, than NGC\,6791, and even more with respect to NGC\,6819 \citep{vandenberg2013}.

On the other hand, cored profiles can be a sign of the existence of massive remnants in the cluster centres, such as stellar mass black holes \citep[e.g][]{merritt2004}. Their presence is able to prevent the segregation of mass of the host systems in their central regions \citep[e.g.][]{pauten2016}. Initially, the formation of black holes in open clusters was discarded because of their low densities. However, recent dynamical studies have demonstrated that binary black holes can form in open clusters by different mechanisms \citep[e.g][]{mapelli2016,kumamoto2020}. In fact, very recently, it has been suggested by \citet{torniamenti2023} the existence of a black-hole population to properly reproduce the density profile of the Hyades open cluster.

\section{Summary and conclusions}\label{sec:conclusions}

In this paper, we determine membership probabilities from \gaia astrometry for stars in the field of view of six of the oldest open clusters in the Milky Way in order to investigate their longevity: NGC\,188, NGC\,2420, NGC\,2425, NGC\,2682, NGC\,6791, and NGC\,6819.

From the study of their spatial distributions, we find that NGC\,188 shows signs of a tail. NGC\,2425 exhibits a non-isotropic distribution of the external region. In the case of NGC\,6819, we notice an elongation of its central region. These structures are aligned with the directions of motion of each cluster.

The derived radial density profiles show some hint of flattening in the outskirts for two of the clusters, NGC\,188 and NGC\,2420. For the other, we only observe a change in the slope. There are also significant differences in the central regions. While NGC\,2682 shows a power-law density increase, the so-called cusp profile, NGC\,6791 exhibits a flat one, known as core profile. 

We use \limepy~and \spes~set of models to characterise the observed radial density profiles. A fraction of potential escapers, stars with enough energy to leave the system or already out of it, is needed to properly reproduce the external regions. These models are used to determine some properties of the clusters such as their current masses, or half-mass, tidal and core radii. We found that the studied clusters are more extended than previously reported in the literature. 

We discuss the obtained results, taking also into account the influence of the Galaxy. A high initial mass is needed to survive but also the location in the Galaxy, and mainly being at a large distance from the Galactic plane, play a role in the clusters' longevity. The most dynamical evolved systems do not fill completely their Roche volumes and lower r$_{\rm t}$/r$_{\rm c}$ ratio. Moreover, all the cluster are segregated in mass but the most dynamically evolved clusters show a lower mass segregation ratio, which should be explained by the fact that these systems have lost a significant fraction of their less massive members.

Finally, the observed cusp-core dichotomy in the central regions may be related to different dynamical evolution. As in the case of globular clusters, cusp profiles may be related with more complex dynamical evolutions with a large presence of binaries and X-ray emission. On the other side, the presence of massive remnants, such as black holes, would prevent the mass segregation and formation of cusp profiles. More data, mainly kinematics, is needed to check these hypotheses. This will provide by \gaia for the brightest targets and by the massive spectroscopic surveys for the fainter ones.

\begin{acknowledgements}
We acknowledge the anonymous referee for his/her very constructive comments, which have contributes to increase the quality of this paper. NAB acknowledge the RECA (Red de Estudiantes Colombianos en Astronom\'{\i}a - Network of Colombian astronomy students) internship programme for the opportunity to begin this work. HT was funded under the Erasmus+ 2020 programme of the European Commission.
This work was supported by the MINECO (Spanish Ministry of Economy, Industry and Competitiveness) through grant ESP2016-80079-C2-1-R (MINECO/FEDER, UE) and by the Spanish MICIN/AEI/10.13039/501100011033 and by "ERDF A way of making Europe" by the “European Union” through grants RTI2018-095076-B-C21 and PID2021-122842OB-C21, and the Institute of Cosmos Sciences University of Barcelona (ICCUB, Unidad de Excelencia ’Mar\'{\i}a de Maeztu’) through grant CEX2019-000918-M.

This work has made use of data from the European Space Agency (ESA) mission
{\it Gaia} (\url{https://www.cosmos.esa.int/gaia}), processed by the {\it Gaia}
Data Processing and Analysis Consortium (DPAC,
\url{https://www.cosmos.esa.int/web/gaia/dpac/consortium}). Funding for the DPAC
has been provided by national institutions, in particular the institutions
participating in the {\it Gaia} Multilateral Agreement.
\end{acknowledgements}

\bibliographystyle{aa} % style aa.bst
\bibliography{long_occ}

\appendix

\section{Membership probability threshold impact}\label{sect:pthreshold}
\begin{table}
\setlength{\tabcolsep}{1mm}
	\centering
	\caption{Best \spes~models fitting parameters for different membership probabilities thresholds}
	\label{tab:test_prob}
	\begin{tabular}{lcccccccccccc}
		\hline
  	Parameter & \multicolumn{4}{c}{NGC\_188} & \multicolumn{4}{c}{NGC\_2420}\\
  & 0.4 & 0.5 & 0.6 & 0.8 & 0.4 & 0.5 & 0.6 & 0.8 \\
		\hline
  $W_0$ & 4.7 & 4.8 & 5.2 & 5.3 & 6.4 & 7.0 & 7.0 & 6.7 \\
  $B$ & 0.10 & 0.13 & 0.10 & 0.09 & 0.0 & 0.0 & 0.0 & 0.0 \\
  $\eta$ & 0.62 & 0.60 & 0.55 & 0.46 & 0.68 & 0.59 & 0.55 & 0.56 \\
  $M_{\rm cl}$ [10$^3$M$_{\odot}$] & 6.8 & 6.7 & 6.7 & 6.3 & 5.0 & 5.0 & 4.5 & 3.5  \\
  $r_{\rm h}$ [pc] & 6.2 & 6.1 & 6.2 & 5.8 & 7.7 & 8.1 & 7.8 & 6.9 \\
 $r_{\rm c}$  [pc]& 3.2 & 3.2 & 3.0 & 3.0 & 2.4 & 2.2 & 2.2 & 2.1  \\
 $r_{\rm t}$  [pc]& 17.2 & 17.3 & 19.1 & 18.4 & 25.0 & 29.0 & 29.3 & 25.1  \\
 fpe\tablefootmark{a} & 0.15 & 0.14 & 010 & 0.05 & 0.18 & 0.12 & 0.09 & 0.09  \\
\hline
Parameter  & \multicolumn{4}{c}{NGC\_2425} & \multicolumn{4}{c}{NGC\_2682}  \\
  & 0.4 & 0.5 & 0.6 & 0.8 & 0.4 & 0.5 & 0.6 & 0.8 \\
\hline
		\hline
  $W_0$ & 7.0 & 6.9 & 7.0 & 6.9 & 6.1 & 6.1 & 6.2 & 6.4 \\
  $B$ & 0.02 & 0.01 & 0.03 & 0.02 & 0.00 & 0.02 & 0.02 & 0.04 \\
  $\eta$ & 0.74 & 0.69 & 0.64 & 0.60 & 0.65 & 0.63 & 0.60 & 0.55 \\
  $M_{\rm cl}$ [10$^3$M$_{\odot}$] & 3.0 & 2.7 & 2.5 & 1.8 & 9.4 & 9.3 & 9.1 & 9.0  \\
  $r_{\rm h}$ [pc] &9.6 & 8.5 & 8.7 & 7.5& 8.2 & 8.2 & 8.1 & 8.1 \\
 $r_{\rm c}$  [pc] & 2.1 & 2.1 & 2.1 & 2.1 & 2.9 & 2.9 & 2.9 & 2.9 \\
 $r_{\rm t}$  [pc]& 30.5 & 28.3 & 30.6 & 27.0 & 17.2 & 17.3 & 19.1 & 18.4 \\
 fpe\tablefootmark{a} & 0.23 & 0.19 & 0.16 & 0.13 & 0.15 & 0.14 & 012 & 0.09  \\
\hline
Parameter & \multicolumn{4}{c}{NGC\_6791}  & \multicolumn{4}{c}{NGC\_6819}\\
& 0.4 & 0.5 & 0.6 & 0.8 & 0.4 & 0.5 & 0.6 & 0.8\\
\hline
\hline
$W_0$& 4.7 & 4.8 & 4.9 & 5.1 & 6.2 & 6.6 & 6.5 & 6.5\\
$B$ & 0.06 & 0.02 & 0.02 & 0.8 & 0.10 & 0.07 & 0.06 & 0.03 \\
$\eta$ & 0.65 & 0.64 & 0.62 & 0.57 & 0.63 & 0.56 & 0.55 & 0.52\\
$M_{\rm cl}$ [10$^3$M$_{\odot}$]& 20.5 & 20.5 & 20.5 & 20.3 & 19.0 & 18.1 & 16.5 & 12.8\\
$r_{\rm h}$ [pc]& 7.9 & 7.9 & 8.0 & 8.3 &9.3 & 9.2 & 8.8 & 8.0\\
$r_{\rm c}$  [pc]& 4.1 & 4.0 & 4.0 & 4.0 & 3.1 & 3.0 & 3.0 & 2.8 \\
$r_{\rm t}$  [pc]& 21.1 & 21.6 & 22.4 & 25.0 & 31.4 & 33.4 & 31.6 & 28.9 \\
fpe\tablefootmark{a}& 0.17 & 0.16 & 0.14 & 0.11 & 0.15 & 0.10 & 0.09 & 0.07\\
\hline
\hline
	\end{tabular}
 \tablefoot{\tablefoottext{a}{Fraction of potential escapers.}}
\end{table}

In this work, we assumed an astrometric membership probability threshold of $p>$0.4 on the basis of the analysis performed by \citet{Soub18} and \citet{ApogCarrera19}. Briefly, these tests are based in comparing the average radial velocity and standard deviation values for several clusters assuming different probability thresholds. The obtained values are stable until a certain value, $p$=0.4, and they began to diverge for lower values. However, it is important to check how this choice affect the results obtained in this paper.

To investigate the impact of this selection, we have obtained the radial density profiles of the six studied clusters assuming different values of $p$: 0.4, 0.5, 0.6 and 0.8. The obtained profiles have been fitted with the \spes~models using the same procedure described in sect.~\ref{sect:model_fit}. The fitted parameters are listed in Table~\ref{tab:test_prob}. There are no significant differences for the \spes~models' parameter values of membership threshold up to $p$=0.6. Only in the extreme case of $p$=0.8 we observe significant variations of the derived parameters for \spes~models. For example, the fraction of potential escapers decreases in about 50\,\% between the case of $p$=0.4 and $p$=0.8.

An extreme membership probability threshold of $p$=0.8 is not reasonable, taken into account the uncertainties. In fact, this value would remove a significant number of stars with radial velocities compatible with that of the clusters according to the test performed by \citet{Soub18} and \citet{ApogCarrera19}, respectively. On the other hand, there are no significant differences between $p$=0.4 and 0.5, respectively.

\section{Impact of contaminants}\label{sec:contaminants}

Another source of uncertainty in our results is the effect of contaminants in the derived radial density profiles and, therefore, in the best model fitted. In fact, in Sect.~\ref{sec:membership} we showed that among the brightest members of the studied cluster could be between 10 and 15\,\% of objects with discrepant radial velocities from \gaia. In order to gauge the impact of potential contaminants in our results, we apply a bootstrap sampling \citet{bootstrap}. Briefly, a bootstrap sample, with the same elements as the initial one, is formed by randomly drawing elements from the initial set without taking any account of whether a point has already been selected or not. This means that any data point may occur no times, one times or many times in any bootstrap sample. From each bootstrap sample, we construct the corresponding radial density profile, which is fitted with the \spes~models. Owing to the amount of computational time required, we used a simply non-linear least square fitting algorithm within the \texttt{LMFIT} \python~ implementation instead of the Markov chain Monte Carlo fitting technique described in Sect.~\ref{sect:model_fit}. This procedure has been repeated a thousand times.

\begin{figure}
	\includegraphics[width=\columnwidth]{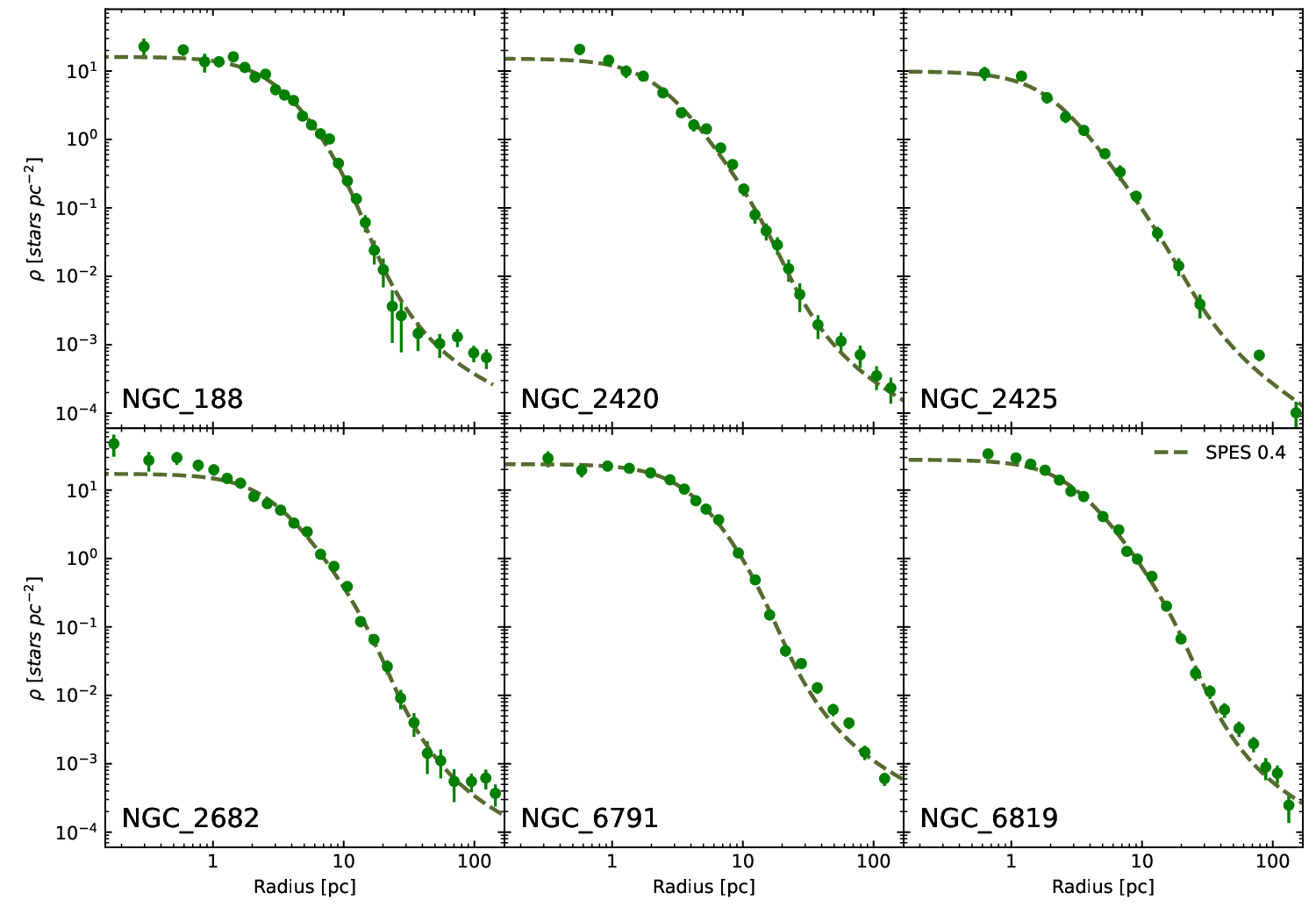}
    \caption{As Fig.~\ref{fig:model_fits} but for radial density profiles derived applying a bootstrap sampling. Dashed lines }
    \label{fig:rdp_bootstrap}
\end{figure}

Obtained results are listed in Table~\ref{tab:spesmodels_bootstrap} and showed in Fig.~\ref{fig:rdp_bootstrap}. The obtained values are in very good agreement with the results described in Sect.~\ref{subsec:results}. Moreover, this analysis provides an alternative check of the robustness of our results.
\begin{table*}
\setlength{\tabcolsep}{1.5mm}
	\centering
	\caption{Best-fitting parameters of \spes~models using a bootstrap sampling}
	\label{tab:spesmodels_bootstrap}
	\begin{tabular}{lcccccccc}
		\hline
		Cluster & $W_0$ & $B$ & $\eta$ & $M_{\rm cl}$ & $r_{\rm h}$ & $r_{\rm c}$& $r_{\rm t}$ & fpe\tablefootmark{a} \\
  & & & & [10$^3$M$_{\odot}$] & [pc] & [pc] & [pc] &\\
		\hline
  NGC\_188 & 4.7$\pm$0.5 & 0.03$\pm$0.40 & 6.7$\pm$0.3 & 0.62$\pm$0.14 & 6.2$\pm$0.3 &    3.1$\pm$0.3 &17.2$\pm$1.5 &0.15$\pm$0.02\\
  NGC\_2420 & 6.4$\pm$0.5 & 0.01$\pm$0.01 & 4.9$\pm$0.4 & 0.66$\pm$0.03 & 7.6$\pm$0.9 &   2.4$\pm$0.2 &25.0$\pm$4.2 &0.17$\pm$0.02\\
  NGC\_2425 & 6.6$\pm$0.6 & 0.01$\pm$0.01 & 2.8$\pm$0.4 & 0.73$\pm$0.05 & 8.3$\pm$1.6 &   2.1$\pm$0.2 &26.9$\pm$5.9 &0.22$\pm$0.03\\
  NGC\_2682 & 6.0$\pm$0.4 & 0.01$\pm$0.01 & 9.3$\pm$0.5 & 0.64$\pm$0.02 & 8.2$\pm$0.5 &   2.9$\pm$0.3 &26.6$\pm$2.9 &0.14$\pm$0.02\\
  NGC\_6791 & 4.3$\pm$0.7 & 0.57$\pm$0.36 & 20.8$\pm$1.5 & 0.61$\pm$0.06 & 8.0$\pm$0.6 &  4.3$\pm$0.5 &22.2$\pm$4.1 &0.16$\pm$0.02\\
  NGC\_6819 & 6.3$\pm$0.3 & 0.01$\pm$0.01 & 18.6$\pm$0.9 & 0.63$\pm$0.02 & 9.2$\pm$0.6 &  3.0$\pm$0.2 &31.1$\pm$3.7 &0.14$\pm$0.01\\
\hline
	\end{tabular}
 \tablefoot{\tablefoottext{a}{Fraction of potential escapers.}}
\end{table*}

\section{Individual model fits} 
\label{appendix:a} 
\begin{figure}[h!]
   \centering
   \includegraphics[width=\columnwidth]{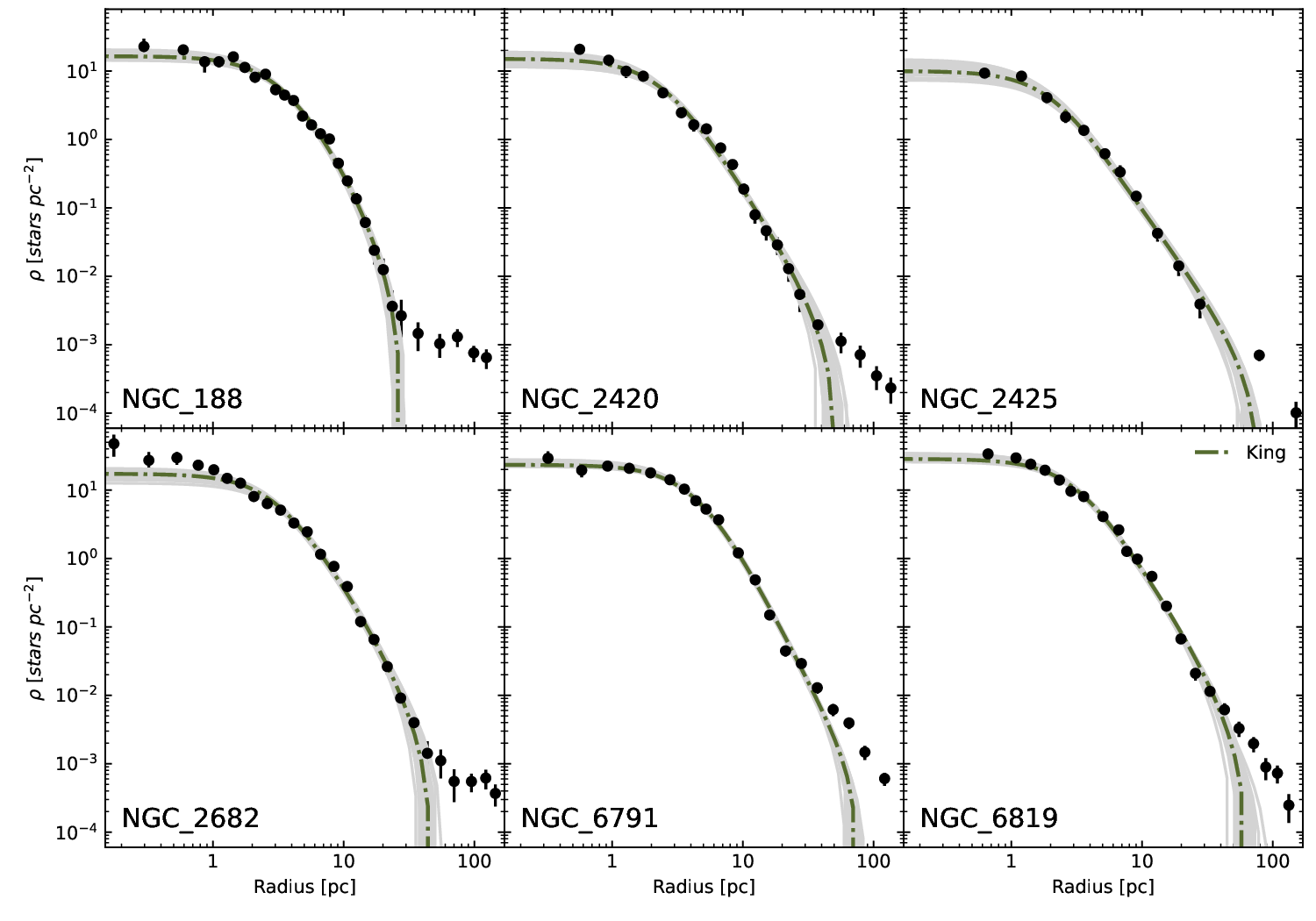}
   \caption{As Fig.~\ref{fig:model_fits} but with the best \limepy~King models fits.}
   \label{fig:king}
\end{figure}

\begin{figure}[h!]
   \centering
   \includegraphics[width=\columnwidth]{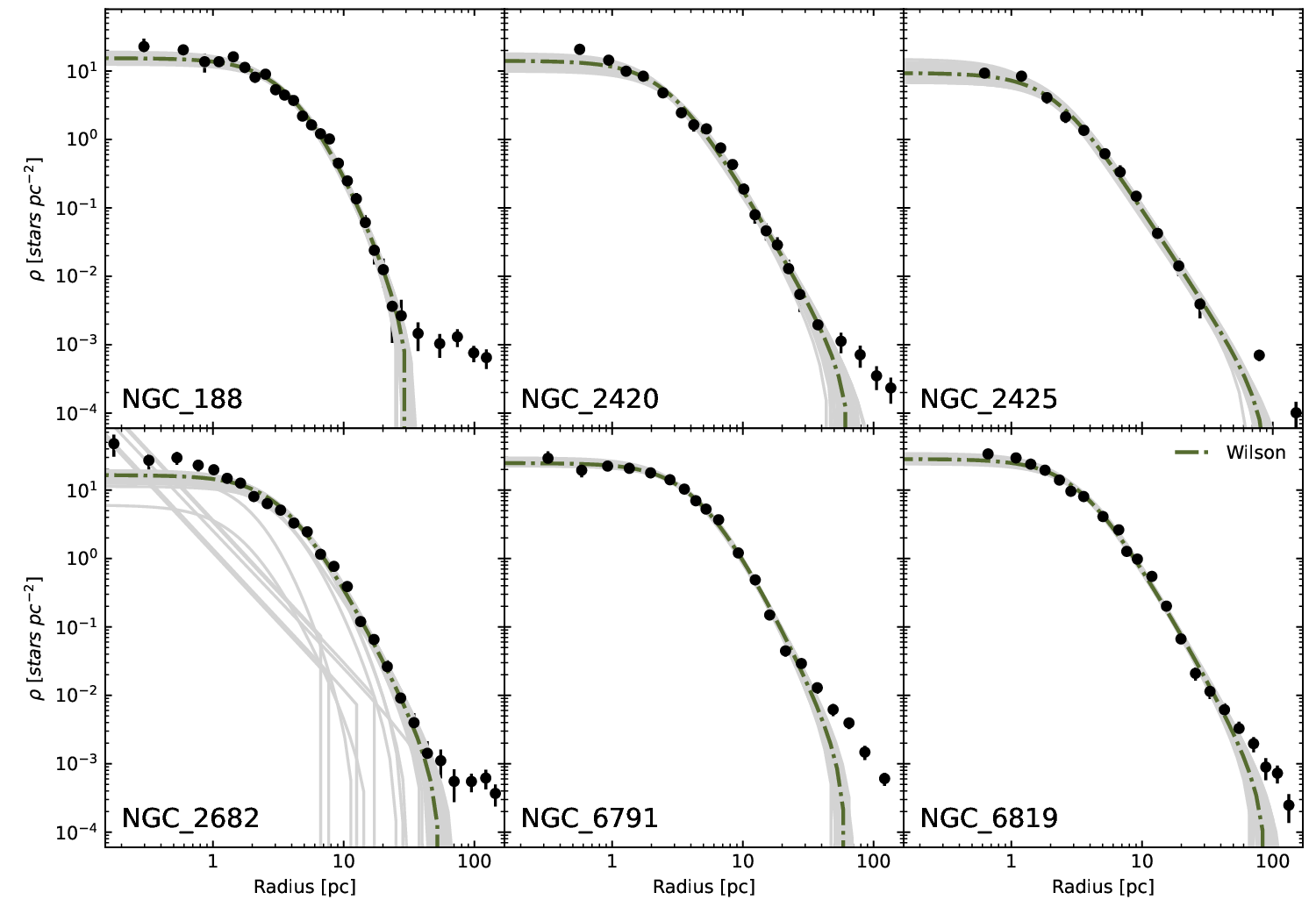}
   \caption{As Fig.~\ref{fig:model_fits} but with the best \limepy~Wilson models fits. Grey areas represent 50, randomly selected, solutions of the 500 total realisations.}
   \label{fig:wilson}
\end{figure}

\begin{figure}[h!]
   \centering
   \includegraphics[width=\columnwidth]{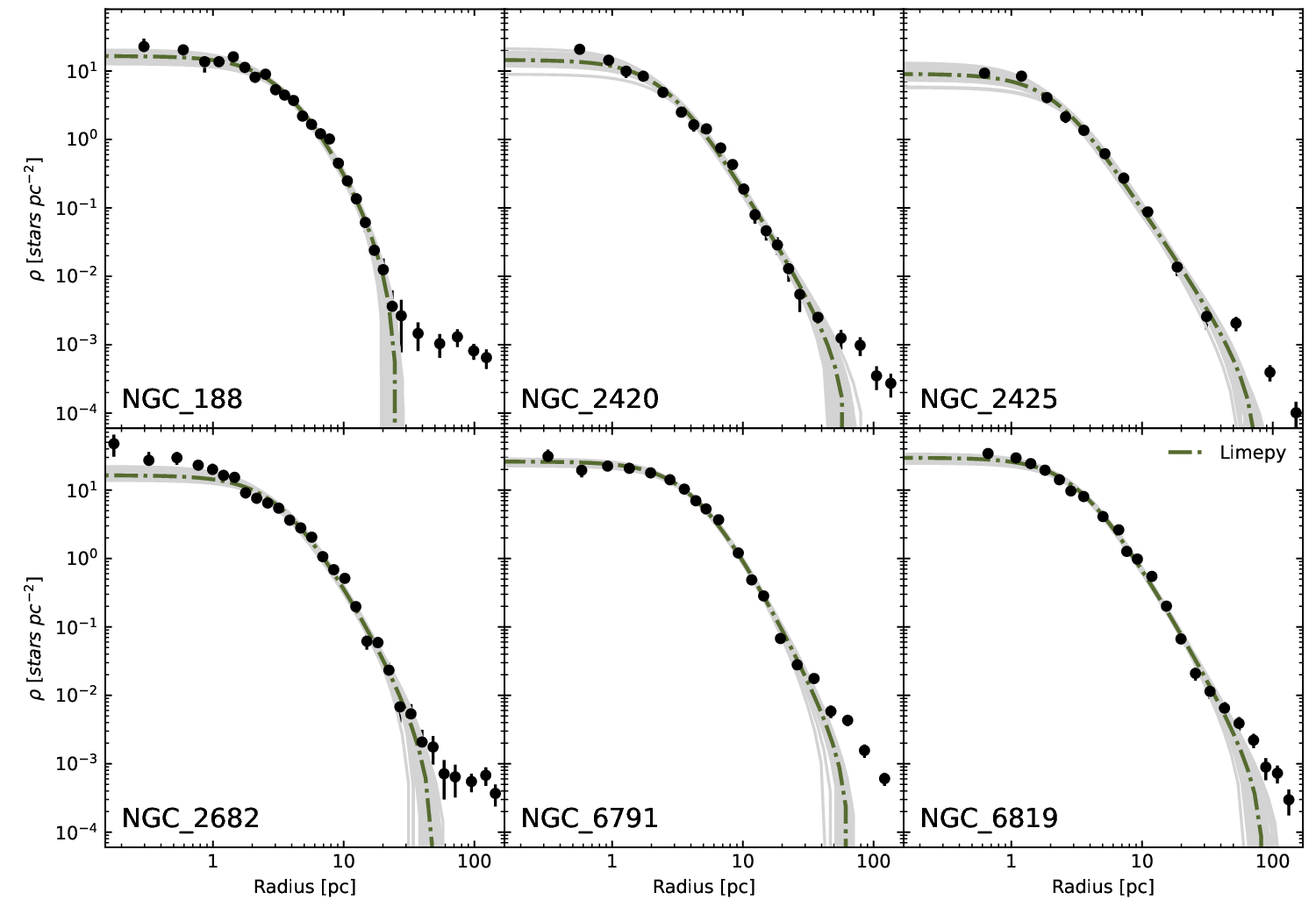}
   \caption{As Fig.~\ref{fig:model_fits} but with the best \limepy~models fits.}
   \label{fig:limepy}
\end{figure}

\begin{figure}[h!]
   \centering
   \includegraphics[width=\columnwidth]{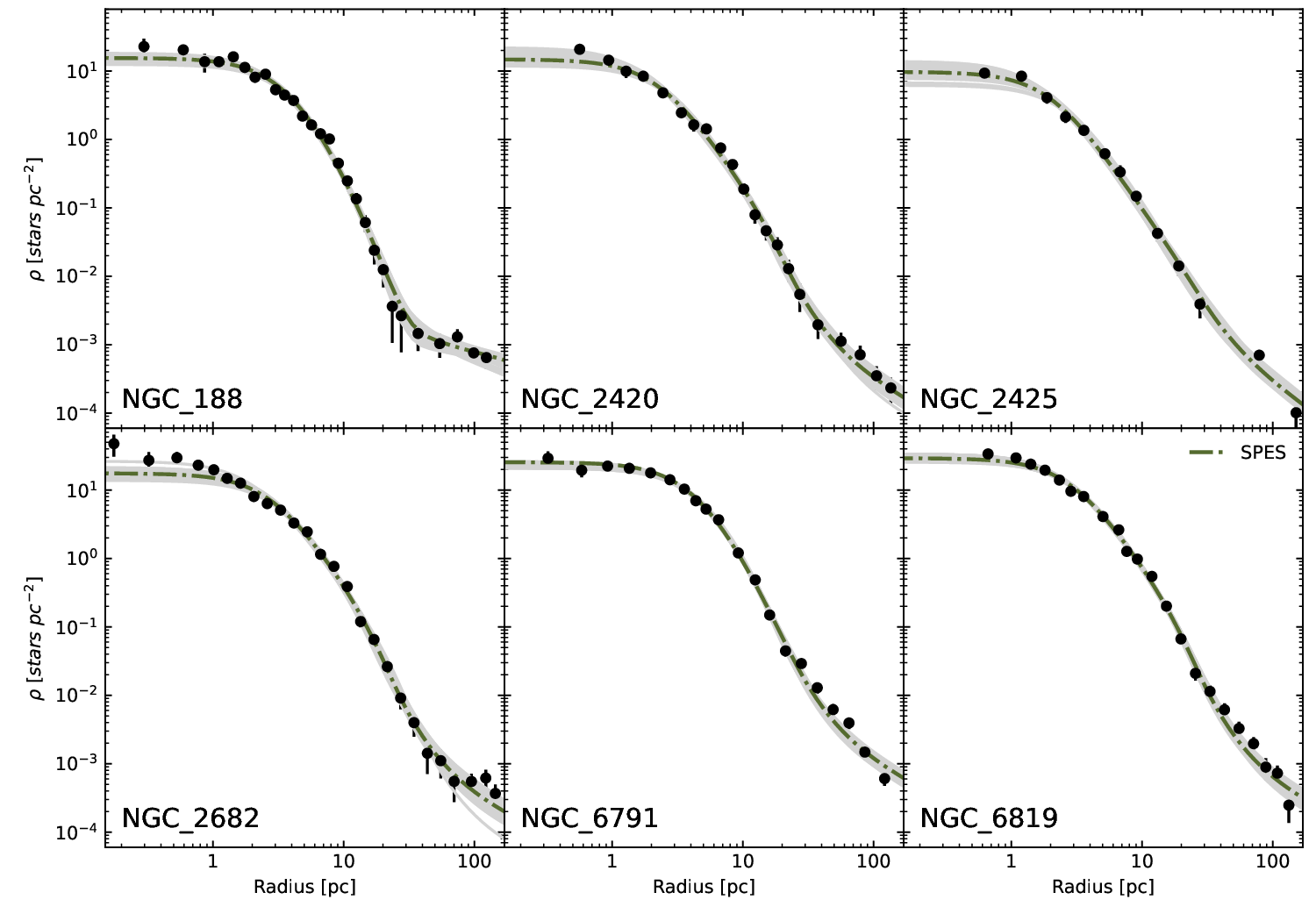}
   \caption{As Fig.~\ref{fig:model_fits} but with the best \spes~models fits.}
   \label{fig:spes}
\end{figure}

\end{document}